\documentclass[api, phf,16pt,showpacs, reprint,onecolumn]{revtex4-1}

\usepackage{amssymb,amsfonts,latexsym,mathtools} %
\usepackage{graphicx,setspace}
\usepackage{graphicx} %
\usepackage{amsmath,euscript,mathrsfs}
\usepackage[hidelinks]{hyperref} % Hyperlinks for equation numbers
\usepackage[usenames,dvipsnames,svgnames]{xcolor}

\hypersetup{
    colorlinks,
    linkcolor={blue!50!black},
    citecolor={blue!50!black},
    urlcolor={blue!50!black}
}

\graphicspath{{Figures/}}

\makeatletter
\def\blfootnote{\gdef\@thefnmark{}\@footnotetext}
\makeatother

\usepackage[all]{hypcap}

\usepackage{array}
\newcolumntype{C}[1]{>{\centering\arraybackslash}m{#1}}

\usepackage[super]{nth}

\usepackage{comment}

\allowdisplaybreaks[1]

\newcommand{\nc}{\newcommand}
\nc{\numberthis}{\addtocounter{equation}{1}\tag{\theequation}}
\nc{\be}{\begin{equation}} \nc{\ee}{\end{equation}}
\nc{\bes}{\begin{equation*}} \nc{\ees}{\end{equation*}}
\nc{\eps}{\varepsilon} \nc{\prt}{\partial} \nc{\ds}{\displaystyle}
\nc{\dsp}[1]{^\mathrm{#1}} \nc{\lb}{\left (} \nc{\rb}{\right )}
\nc{\lset}{\left \{} \nc{\rset}{\right \}} \nc{\eqtext}[1]{\quad
\text{#1} \quad} \nc{\lsq}{\left [} \nc{\rsq}{\right ]}
\nc{\half}{\frac{1}{2}} \nc{\RA}{\quad \Rightarrow \quad}
\nc{\coshn}[2]{\mathop{\rm cosh}\nolimits ^{#1} \lb #2 \rb}
\nc{\sechn}[2]{\mathop{\rm sech}\nolimits ^{#1} \lb #2 \rb}
\nc{\tanhn}[2]{\mathop{\rm tanh}\nolimits ^{#1} \lb #2 \rb}
\nc{\arccosh}[1]{\mathop{\rm arccosh}\nolimits \lb #1 \rb}
\nc{\ob}[1]{\overbrace{#1}} \nc{\ub}[1]{\underbrace{#1}}
\nc{\field}[1]{\mathbb{#1}} \nc{\inflim}{_{-\infty}^{\infty}}
\nc{\dd}[1]{\; \mathrm{d} #1} \nc{\diff}[2]{\frac{\mathrm{d}
#1}{\mathrm{d} #2}} \nc{\diffn}[3]{\dfrac{\mathrm{d}^{#1}
#2}{\mathrm{d} #3^{#1}}} \nc{\pdd}[1]{\; \partial #1}
\nc{\pdiff}[2]{\frac{\partial #1}{\partial #2}}
\nc{\pdiffn}[3]{\dfrac{\partial^{#1} #2}{\partial #3^{#1}}}
\nc{\nt}{\newtheorem} \nc{\ntc}{\newtheorem*}
\nt{theorem}{Theorem}[section]
\nt{definition}{Definition}[section] \nt{exa}{Example}[section]
\nt{lem}{Lemma}[section] \nt{cor}{Corollary}[section]
\nt{pro}{Proposition}[section]

\newcommand{\sech}{\mathop{\rm sech}\nolimits}
\newcommand{\vp}{\varphi}%

\begin{document}

\title{\vspace*{1cm}
Soliton spectra of random water waves in shallow basins} %

\author{J.-P.~Giovanangeli$^{1)}$, C.~Kharif$^{1)}$, and Y.A. Stepanyants$^{2, 3)}$
\blfootnote{Corresponding author: Yury.Stepanyants@usq.edu.au}\\}

\affiliation{\vspace*{0.5cm} $^{1)}$ Aix Marseille Universit\'{e},
CNRS, Centrale
Marseille, IRPHE UMR 7342, 13384, Marseille, France; \\%
$^{2)}$ Faculty of Health, Engineering and Sciences, University of
Southern Queensland, \\Toowoomba, QLD, 4350, Australia and \\
$^{3)}$ Department of Applied Mathematics, Nizhny Novgorod State
Technical University n.a. R.E. Alekseev, \\Nizhny Novgorod,
603950, Russia.}

%\date{\today}

\begin{abstract}%
\vspace*{1cm} %
Interpretation of random wave field on a shallow water in terms of
Fourier spectra is not adequate, when wave amplitudes are not
infinitesimally small. A nonlinearity of wave fields leads to the
harmonic interactions and random variation of Fourier spectra. As
has been shown by Osborne and his co-authors, a more adequate
analysis can be performed in terms of nonlinear modes representing
cnoidal waves; a spectrum of such modes remains unchanged even in
the process of nonlinear mode interactions. Here we show that
there is an alternative and more simple analysis of random wave
fields on shallow water, which can be presented in terms of
interacting Korteweg--de Vries solitons. The data processing of
random wave field is developed on the basis of inverse scattering
method. The soliton component obscured in a random wave field is
determined and a corresponding distribution function of number of
solitons on their amplitudes is constructed. The approach
developed is illustrated by means of artificially generated
quasi-random wave field and applied to the real data
interpretation of wind waves generated in the laboratory wind
tank. \\



\end{abstract}

%\pacs{47.35.Bb, 47.35.-i, 47.35.Fg}
%47.35.Pq, 68.03.Cd,47.15.ki}

\maketitle

\vfill

\clearpage

\section{Introduction}

The traditional approach in the problem of wind wave study is
based on the analysis of Fourier spectra and determination of
their peculiarities. There is a vast number of papers both
theoretical and experimental where this problem has been
considered; it is impossible to list all of them here. Therefore,
we refer only to the review chapter ``Wind waves'' by Zaslavsky
and Monin in the book \cite{KamMon} where a reader can find key
references in this field. A lot of interesting and useful
information has been obtained about wind waves, and their analysis
and interpretation have been implemented in terms of Fourier
spectra.

Meanwhile, the Fourier analysis of wind waves provides researchers
with only some piece of objective information whereas many
important features of wind waves remain hidden. One of the serious
obstacles making the Fourier analysis ineffective in application
to surface oceanic waves is the nonlinear character of such waves,
whereas the Fourier analysis is a linear operation applicable to
systems obeying the superposition principle.

Osborne with co-authors (see, e.~g., \cite{Osb1, Osb2, OSBC, SBOP,
Osborne} and references therein) have developed the method of
nonlinear spectral analysis of shallow water waves described by
the Korteweg--de Vries (KdV) equation. Osborne's approach is based
on the application of the inverse scattering method (ISM) to the
analysis of random field data in a one-dimensional space domain
with the periodic boundary conditions. The main idea of his
approach was in the presentation of complex initial disturbance in
terms of a set of elliptic functions (cnoidal waves). These
functions can be considered as the nonlinear eigenmodes which are
preserved in the process of wave field evolution in contrast to
the linear sinusoidal eigenmodes. This means that a nonlinear wave
spectrum calculated on the basis of these modes is invariant in
time while a usual Fourier spectrum is variable due to nonlinear
interactions between the different sinusoidal harmonics. An
important feature of nonlinear eigenmodes is their reducibility to
the sinusoidal eigenmodes in the case of small amplitudes. In
other words, a nonlinear spectrum naturally reduces to the Fourier
spectrum if the analyzed wave field is quasi-linear. However, the
mathematical and numerical machinery used for calculation of
nonlinear eigenmodes is not simple in contrast to the linear case
and, apparently, it is impractical especially for the applied
mariner engineers.

Here we propose a very similar to Osborne's, but a bit different
approach to the analysis of random water waves that is also based
on the application of ISM. The essential feature of our approach
is the interpretation of a random initial wave field in terms of
an ensemble of solitons and quasi-linear ripples rather than the
set of elliptic eigenmodes (a similar approach was recently
realised in Ref. \cite{Costa}). The idea is illustrated by an
example of shallow water waves described by the classical KdV
equation with the random initial data. If one takes some portion
of random field data (which should be long enough), the number of
solitons, their amplitudes, speeds, characteristic durations,
etc., can be calculated then by means of ISM (e.g., by solving
numerically the associated Sturm--Louiville problem) or by the
direct numerical simulation of the corresponding KdV equation. In
both cases, the numerical codes are currently very well developed
and easily available. In the meantime, the knowledge of number of
solitons obscured in the random wave field, their parameters and
statistics is a matter of independent interest per se. We describe
our approach below in detail and give some examples (preliminary
results were reported at the conference OCEANS'13 MTS/IEEE in San
Diego, USA \cite{GKRS-13}).

Before we start, it is useful to remind that the soliton
turbulence of rarified soliton ensembles in strongly integrable
systems is trivial to certain extent -- the distribution function
of solitons is unchanged in time \cite{Zakh-71, Zakh-09} (the
definition of strongly and weakly integrable systems is given in
\cite{Zakh-09}). This is a consequence of trivial character of
soliton interactions in such systems. The solitons do not change
their parameters after collisions, and only the paired collisions
occur between them. However, the dynamics of a dense ensemble of
solitons is more complicated and soliton turbulence is nontrivial
even in the integrable nonlinear wave equations \cite{El-03,
ElKam-05, El-16}. In particular, in Ref. \cite{El-16} the
quantitative criterion of the term ``dense soliton gas'' was
introduced and was shown that the density of KdV soliton gas is
bounded from above. The critical gas density, apparently, depends
on soliton distribution function, which makes important the
determination of such function in a concrete physical problem such
as water wave turbulence. Numerical experiments confirming the
developed theory in Ref. \cite{El-16} for the particular model
distribution functions were reported in \cite{Carbone-16}.

The KdV model considered here is the typical example of strongly
integrable system applicable to real physical systems. This makes
topical the development of handy methods of extraction of soliton
distribution function from natural complex fields. One of the
experimental approaches to the solution of this practical problem
has been considered for internal waves in Okhotsk Sea
\cite{Nagovitsyn} and another example of processing of surface
wave observational data was reported in Ref. \cite{Costa}. We
hope, this publication will stimulate further interest to this
important problem.

\section{The Korteweg--de Vries model and data processing}
\label{Sect2.1}

\begin{figure}[h!]%
\vspace*{-10cm}%
\centerline{\includegraphics[width=16cm]{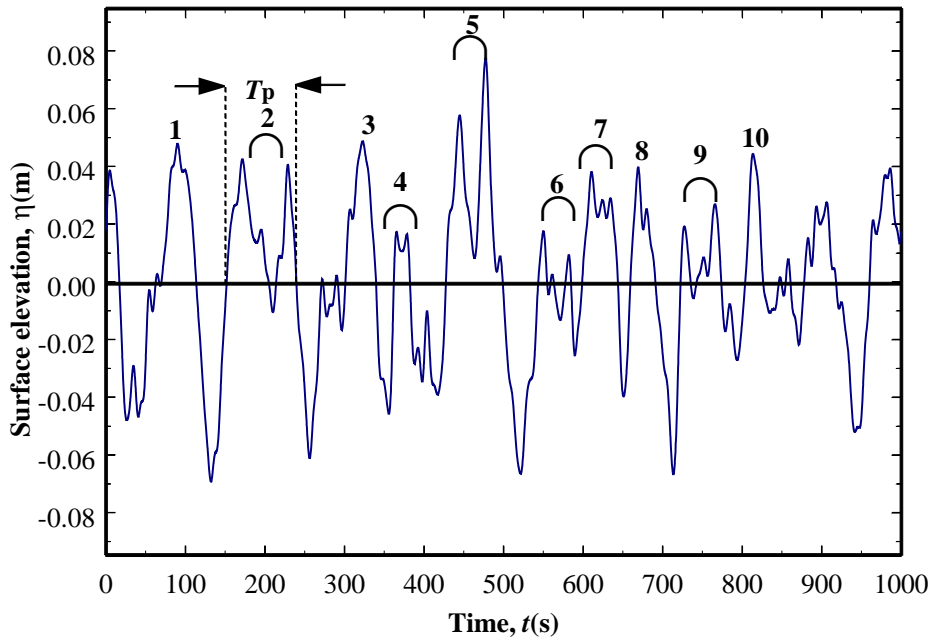}} %
\vspace*{-2cm}%
\caption{A typical example of random surface waves in a shallow
water basin.}
\label{f01} %
\end{figure}

Let us assume that there is a data of recorded surface waves at
some fixed point $x_0$ of a shallow-water basin. So that the
elevation $\eta$ of the water level at this point is the known
function of time: $\eta(x_0,t) = f(t)$ where $f(t)$ is some random
function. A typical example of random surface waves usually
measured in shallow-water basin is shown in Fig.~\ref{f01}. Here
these data were artificially produced by means of a computer using
the random number generator just to illustrate the idea of our
approach.

For the description of further space evolution of data measured at
the point $x_0$, the so-called, timelike KdV equation (TKdV Eq.)
\cite{OSBC} is used:
\be %
\label{e01} %
\frac{\partial\eta}{\partial x} + \frac{1}{c_0}
\frac{\partial\eta}{\partial t} - \alpha
\eta\frac{\partial\eta}{\partial t} - \beta
\frac{\partial^3\eta}{\partial t^3} = 0, %
\ee %
where $c_0 = \sqrt{gh\vphantom{h^2}}; \quad \alpha = 3/(2c_0h);
\quad \beta = h^2/(6c_0^3)$ with $g$ being the acceleration due to
gravity and $h$ being an unperturbed water depth.

In the process of evolution of an initial perturbation one can
expect emergence of a number of solitons with different amplitudes
and phases (e.g., time markers of soliton maxima in the wave
record at a given point of observation). Soliton solution to the
TKdV equation (\ref{e01}) has the form:
\be %
\label{e011} %
\eta(x,t) = A\sech^2{\frac{t - x/V}{T}}, %
\ee %
where the velocity $V$ and duration $T$ are relate to the
amplitude $A$:
\be %
\label{e012} %
V = \frac{c_0}{1 - c_0\alpha A/3} \approx c_0\left(1 +
\frac{c_0\alpha A}{3}\right), \qquad T =
\sqrt{\frac{12\beta}{\alpha A}}, %
\ee %
the approximate formula is valid for small amplitude solitons when
$c_0\alpha A/3 \ll 1$.

\bigskip
%\newpage

As it was mentioned in \cite{OSBC}, ``it may not be possible to
observe solitons in real space'' because of numerous nonlinear
interactions of solitons both with each other and with chaotic
radiation background component. However, ``it would be naive to
conclude that solitons are not present or that their dynamics are
not important'' in the evolution of the initial perturbation.
Inasmuch as solitons are very stable with respect to interaction
with others wave perturbations and influence of external effects
(such as viscosity, inhomogeneity, etc), it is a matter of
interest to extract them from the irregular components of a wave
field and to describe their statistical properties and
contribution to the total wave energy.

The solution of this problem can be done by the following way. Let
us consider a very long portion of recorded measurement data of
surface perturbation at any given point $x_0$. The characteristic
duration of this portion $T_p$ is assumed to be much greater than
the typical soliton time scale $T$. Let us represent a
perturbation with the help of some dimensionless function
$\vp(t)$:
\be %
\label{e02} %
\eta(0,t) = U\vp(t/T_p), %
\ee %
where $U$ is the characteristic wave ``amplitude'', e.g., the
maximum value of perturbation $\eta(0,t)$ in the considered
portion of data.

By means of the transformation
\be %
\label{e03} %
u = \frac{\eta}{U}, \qquad \xi = -\frac{\alpha U}{T_p}x, \qquad
\tau = \frac 1{T_p}\left(t -
\frac{x}{c_0}\right), %
\ee %
Eq.~(\ref{e01}) and the corresponding ``initial'' perturbation
(\ref{e02}) can be reduced to the standard form \cite{Karp} (the
term ``initial'' used here as it is traditionally used in
mathematics for the solution of Cauchy problem of differential
equations, but in fact, the perturbation is given at the fixed
spatial point, i.e., it is rather the boundary condition):
\be %
\label{e04} %
u_{\xi} + uu_{\tau} + \frac 1{\sigma^2}u_{\tau\tau\tau} = 0 %
\ee
\be %
\label{e05} %
u(0,\tau) = \vp(\tau) %
\ee %
with one dimensionless parameter $\sigma^2$ known in the
oceanography as the Ursel parameter and defined as
\be %
\label{e06}%
\sigma^2 = \frac{\alpha UT_p^2}{\beta}. %
\ee

As it was mentioned above, we consider the case when the duration
of the perturbation is long enough, so that $\sigma^2 \gg 1$. In
this case the number of solitons obscured in the ``initial''
perturbation is also very big in general, and it is reasonable to
describe them by the distribution function $f(A)$. This function
determines the number of solitons $dN$ within the interval ($A,\,
A+dA$) \cite{Karp}
\be %
\label{e07} %
dN = f(A)dA. %
\ee

According to the theory developed in \cite{Karp}, the distribution
function can be calculated at large values of $\sigma$ by means of
the formula:
\be%
\label{e08} %
f(A) =
\frac{\sigma}{4\pi\sqrt{3U}}\int\limits_L\frac{d\tau}{\sqrt{2U\vp(\tau)
 - A}}, %
\ee %
where the interval of integration $L$ is determined by the
condition
\be %
\label{e09} %
2U\vp(\tau) > A. %
\ee

As follows from Eq.~(\ref{e08}), soliton amplitudes are
distributed in the interval \cite{Karp}
\be %
\label{e10} %
0 < A < 2U\mbox{max}[\vp(\tau)], %
\ee %
and their total number can be found from the formula
\be %
\label{e11} %
N = \int\limits_0^{\infty}f(A)\,dA =
\left(\frac{\sigma}{\pi\sqrt{6}}\right) \int\limits_{\vp(\tau)\;
>\; 0}\sqrt{\vp(\tau)}\,d\tau. %
\ee

Therefore for large $\sigma$ the total number of solitons is
determined only by those intervals of $\tau$-axis where function
$\vp(\tau)$ is nonnegative!

As well known, the KdV equation possesses an infinite number of
conserved densities $I_n$ \cite{Whth, Karp, AblSeg}. One of them,
\be %
\label{e111} %
I_2 = \int\limits_{-\infty}^{+\infty}
\eta^2(x,t)\,dt, %
\ee %
is proportional to wave energy and therefore is of a special
physical interest. For the sake of simplicity we will call this
value simply the ``energy''.

The fraction of energy of a nonsoliton component of a perturbation
to the total energy of ``initial'' perturbation can be determined
by means of formulae (18.27), (18.28) and (19.8) in Ref.
\cite{Karp}:
\be %
\label{e12} %
\frac{I_2^{ns}}{I_2^{tot}} = \frac{\int\limits_{\eta(0,\tau)\; <\;
0}\eta^2(0,\tau)\,d\tau}{\int\limits_0^{T_p}\eta^2(0,\tau)\,d\tau}.%
\ee

The total energy of a soliton component in the wave field can be
readily calculated, if soliton amplitudes are known:
\be %
\label{e13} %
I_2^{sol} = 4\sqrt{\frac{\beta}{3\alpha}}\sum\limits_{k =
1}^{N}A_k^{3/2} = \frac{4h}{3\sqrt{3g}}\sum\limits_{k =
1}^{N}A_k^{3/2}. %
\ee %
In the last expression the values of coefficients $\alpha$ and
$\beta$ for surface water waves were used (see above after
Eq.~(\ref{e01})).

To illustrate the idea of this approach we consider below several
examples of different ``initial'' perturbations.

\subsection{Sinusoidal perturbation and its modifications}

\subsubsection{A sinusoidal perturbation}
\label{Subsec2.1.1}

Let us study the KdV Eq.~(\ref{e01}) with the periodical
conditions in time. Let us set $h = 1$ m, then $c_0 = 3.13$ m/s,
$\alpha = 0.479$ s/m$^2$, $\beta = 0.00543$ s$^3$/m. Assume then
that the perturbation has the form
\be %
\label{e14} %
\eta(0,t) = A\sin{\omega t}, %
\ee %
where $A = 0.01$ m, $\quad \omega = 0.0628$ 1/s.

The value of Ursel parameter for this perturbation is $\sigma^2 =
555 \gg \sigma_s^2$, where $\sigma_s^2 \equiv 12$ is the value of
Ursel parameter for a single soliton regardless of its amplitude
and duration \cite{Karp}.

\begin{figure}[ht]%
\vspace*{-10cm}%
\centerline{\includegraphics[width=16cm]{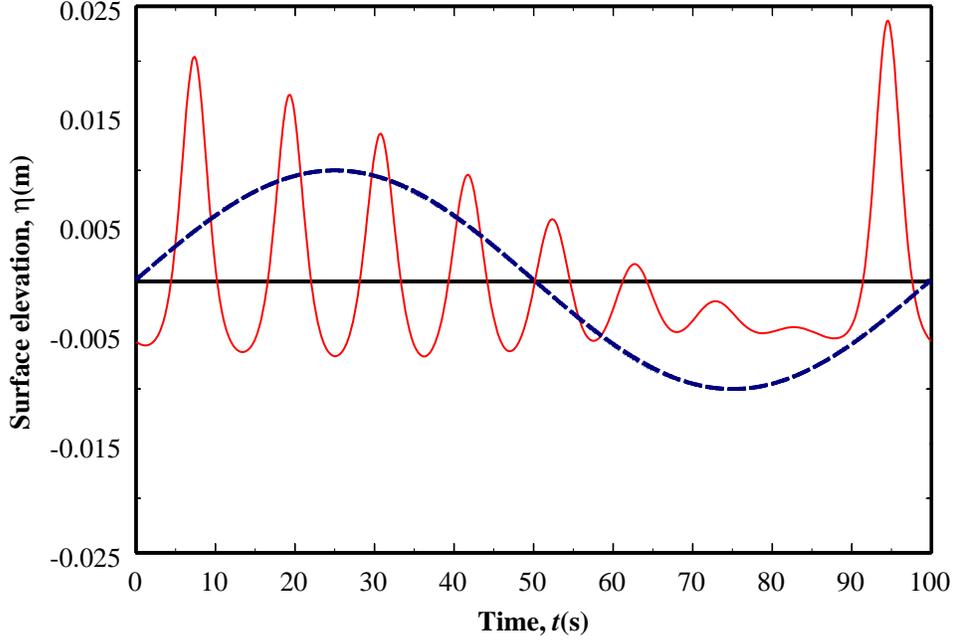}} %
\vspace*{-2cm}%
\caption{(color online) Sinusoidal initial perturbation (dashed
line) and the result of its evolution at the distance $x =
1.1\cdot10^4$ m (solid line).}
\label{f02} %
\end{figure}

Figure \ref{f02} shows this sinusoidal initial perturbation and
the result of its evolution at the distance $x = 1.1\cdot10^3$ m
when it is completely disintegrated into the sequence of solitons.
After that moment, solitons begin interact with each other and the
wave field looks much more complex. In Fig.~\ref{f02} one can
distinguish nine different solitons whose amplitudes are presented
in Table 1. Soliton amplitudes were conditionally defined here as
the difference between their maxima and the mean value of two
nearest minima.

\clearpage

Table 1. Amplitudes of solitons, $A_k$, emerging from the
different initial perturbations. The fifth raw in the Table shows
the relative difference $RD$ in the soliton amplitudes of rows
three and four in percents, this will be explained in subsection
\ref{Subsec2.1.3}.

\smallskip

{\hspace{5mm} %
\begin{tabular}{|l|c|c|c|c|c|c|c|c|c|} %
\hline %
\phantom{wwwww}$A_k \to$ & $A_1$ & $A_2$ & $A_3$ & $A_4$ & $A_5$ & $A_6$ & $A_7$ & $A_8$ & $A_9$ \\
Initial pert. $\downarrow$&(mm)&(mm)&(mm)&(mm)&(mm)&(mm)&(mm)&(mm)&(mm) \\
\hline %
Sin., periodic & $29.39$ & $26.7$ & $23.72$ & $20.37$ & $16.38$ & $11.61$ & $6.625$ & $2.876$ & $0.792$ \\
\hline %
Sin., pulse & $17.77$ & $13.24$ & $8.927$ & $4.743$ & $1.092$ & $$ & $$ & $$ & $$ \\
\hline %
Half-sin., pulse & $17.71$ & $13.24$ & $8.863$ & $4.738$ & $1.171$ & $$ & $$ & $$ & $$ \\
\hline %
$RD\cdot 100\%$ & $0.34$ & $0$ & $0.723$ & $0.1$ & $-6.75$ & $$ & $$ & $$ & $$ \\
\hline %
\end{tabular}}

\bigskip

The total energy of the sinusoidal initial perturbation over a
period $T = 2\pi/\omega$ can be easily evaluated
\be%
\label{e15} %
I_2^{tot} = \frac 12\int\limits_0^{T}\eta^2(0,t)\,dt =
\frac{A^2}{2}\int\limits_0^{T} \sin^2{\omega t}\,dt =
\frac{A^2T}{4} = 2.5\cdot10^{-3}\, \mbox{m}^2\,\mbox{s}. %
\ee

Similarly, the energy of the non-soliton component of the
perturbation is (see Eq.~(\ref{e12}))
\be %
\label{e16} %
I_2^{ns} = \int\limits_{\eta(0,\tau)\; <\;0}\eta^2(0,\tau)\,d\tau
= \frac{A^2}{2}\int\limits_{T/2}^{T} \sin^2{\omega t}\,dt =
\frac{A^2T}{8} = 1.25\cdot10^{-3}\, \mbox{m}^2\,\mbox{s}. %
\ee

Thus, $I_2^{ns} = I_2^{tot}/2$, and the energy of all solitons in
this case should be $I_2^{sol} = I_2^{tot}/2 =
1.25\cdot10^{-3}\,\mbox{m}^2\, \mbox{s}$. Let us calculate however
this energy directly by means of Eq.~(\ref{e13}) using data of
Table 1 for the soliton amplitudes %
\be
\label{e17}%
I_2^{sol} = \frac{4h}{3\sqrt{3g}}\sum\limits_{k = 1}^{9}A_k^{3/2}
= 4.922\cdot 10^{-3}\, \mbox{m}^2\,\mbox{s}. %
\ee

This value about four times greater then the expected. The
discrepancy can be explained by the incorrect counting of number
of solitons, as well as their amplitudes. As has been shown in
\cite{SalMauEng, Salupere-02, Salupere-03}, the problem of
detection of solitons emerging from the harmonic perturbations is
not so simple even within the KdV equation (disintegration of sine
wave onto set of solitons was recently studied also within the
Gardner equation \cite{Kurkina-16A, Kurkina-16B}). Actual number
of solitons is always greater then that at the instant of time
when they appear from the initial perturbation for the first time
in the ordered form. Their amplitudes are also different from
those which are seen in Fig.~\ref{f02}. We will come back to this
issue a bit later, and now let us consider a pulse-type initial
perturbations defined on a compact support. \\

\subsubsection{A pulse of sinusoidal profile}
\label{Subsec2.1.2}

\begin{figure}[ht]
\vspace*{-10cm}%
\centerline{\includegraphics[width=16cm]{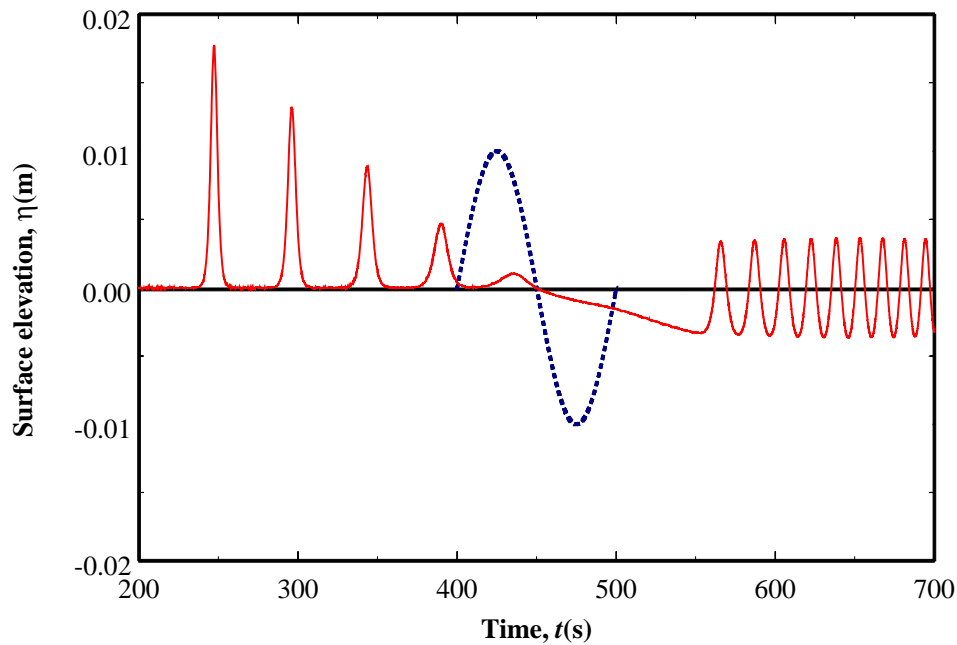}} %
\vspace*{-2cm}%
\caption{(color online) A sinusoidal-shape pulse as the initial
perturbation (dashed line) and the result of its evolution at the
distance $x = 6\cdot10^4$ m (solid line). Only a fragment of a
very long quasi-sinusoidal tail behind five solitons is shown in
the figure.}
\label{f03}%
\end{figure}

Consider now a pulse-type initial perturbation having the shape of
one period of sine with the same amplitude and characteristic
duration as in the previous case (see dashed line in
Fig.~{\ref{f03}}). In the process of evolution this perturbation
disintegrates into a sequence of five solitons (see solid line in
Fig.~{\ref{f03}}), whose amplitudes are indicated in the third row
of Table 1. An intense oscillatory tail behind the solitons is
also appeared; the head portion of this tail is shown in
Fig.~{\ref{f03}}.

By comparison of rows two and three of Table 1, one can see that
the number of solitons and their amplitudes are absolutely
different in the periodic and nonperiodic cases.

As all these solitons are practically independent at the distance
indicated in Fig.~\ref{f03}, their energies can be calculated
independently. The calculation of cumulative energy of these five
solitons yield: %
\be
\label{e18}%
I_2^{sol} = \frac{4h}{3\sqrt{3g}}\sum\limits_{k = 1}^{5}A_k^{3/2}
= 1.2531\cdot 10^{-3}\, \mbox{m}^2\,\mbox{s}. %
\ee

This result already agrees quite well with the theoretical
prediction.

\subsubsection{A half-sine pulse}
\label{Subsec2.1.3}

\begin{figure}[ht]
\vspace*{-10cm}%
\centerline{\includegraphics[width=16cm]{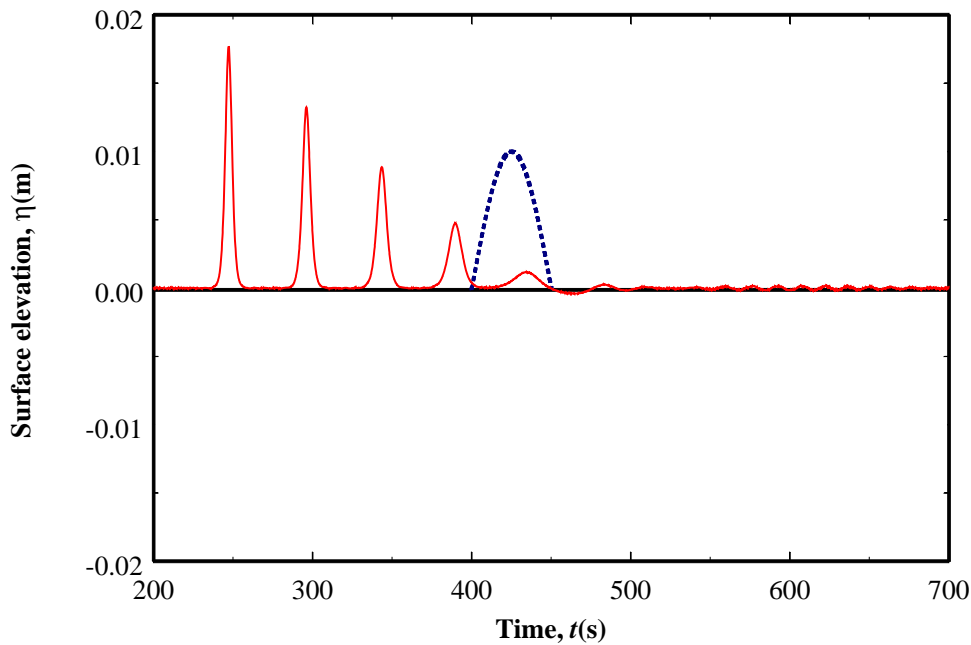}} %
\vspace*{-2cm}%
\caption{(color online) A half-sine pulse as the initial
perturbation (dashed line) and the result of its evolution at the
distance $x = 6\cdot10^4$ m (solid line).}
\label{f04}%
\end{figure}

Let us consider now another pulse-type perturbation which contains
only a half period of a sine-function of a positive polarity (see
dashed line in Fig.~{\ref{f04}}).

This perturbation disintegrates into the same number of solitons
as in subsection \ref{Subsec2.1.2}. At the same distance $x =
6\cdot 10^4$ m, the soliton amplitudes are practically the same as
in the previous case (cf. rows three and four in Table 1). The
fifth raw in Table 1 shows the relative difference in the soliton
amplitudes in percents, $RD = (A_i' - A_i'')/A_i''$, where $A_i'$
and $A_i''$ are the amplitudes of i-th solitons emerged from the
sinusoidal and half-sine pulses respectively.

As one can see from Table 1 and Figs.~\ref{f03} and \ref{f04},
there is no much differences in the corresponding soliton
amplitudes for the sinusoidal and half-sine pulses. In the next
Subsection we will consider a simple analytical example to justify
this numerical observation.

\subsection{Analytical solutions to the associated
Schr\"odinger equation for the rectangular and meander-type
pulses}

To understand better the regularity of distribution of amplitudes
of solitons emerging from initial perturbations, let us consider
two model initial perturbations: the rectangular pulse of positive
polarity and amplitude $U_0$ (Fig.~5a) and meander-type pulse
whose positive part coincides with the above rectangular pulse and
negative part is the same but of opposite polarity (Fig.~5b).
Assume that the duration of positive rectangular pulses are $T$.

\begin{figure}[ht]
%\vspace*{-10cm}%
\centerline{\includegraphics[width=12cm]{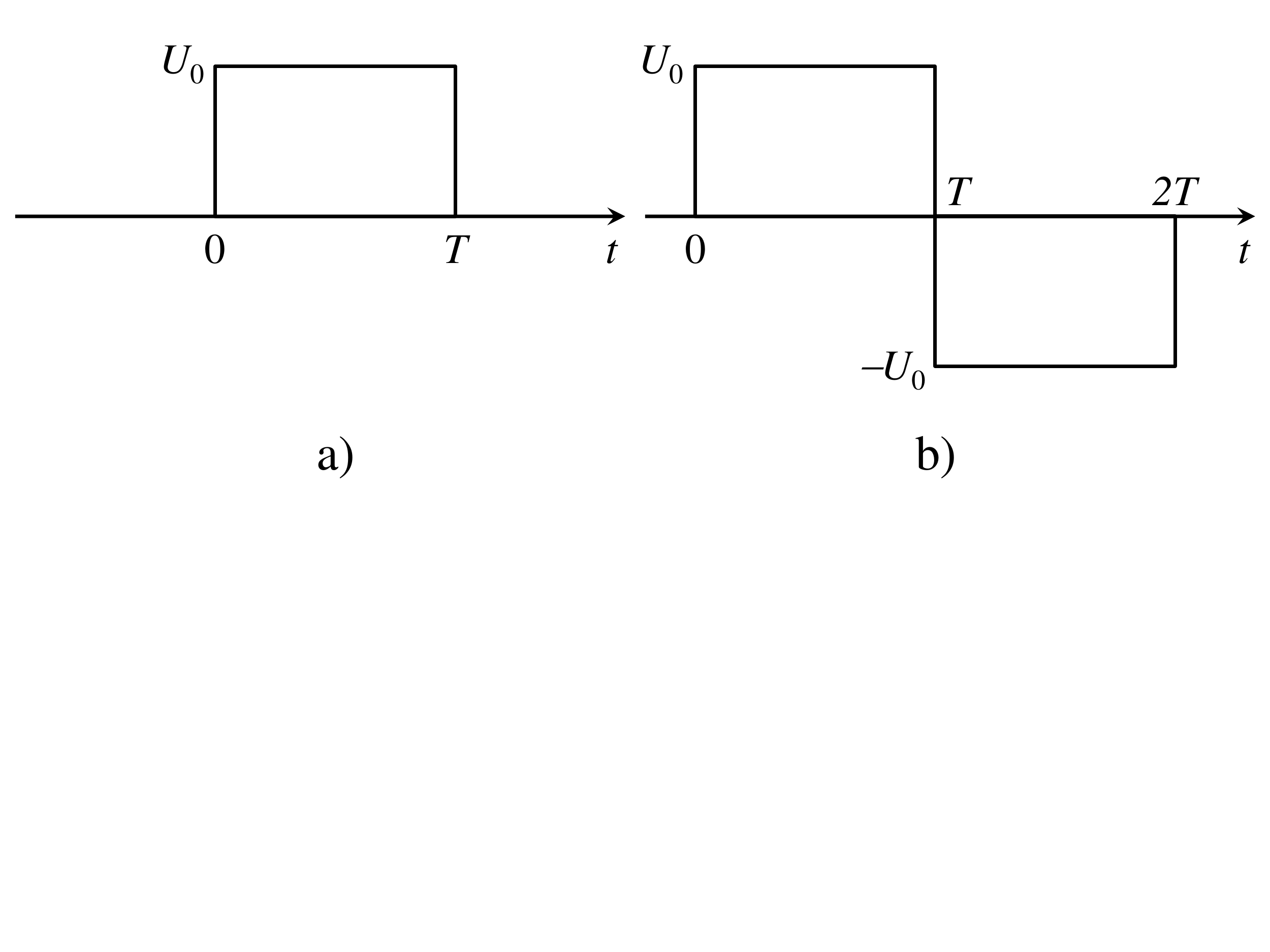}} %
\vspace*{-4.5cm}%
\caption{Two model initial perturbations: the rectangular pulse
(a) and meander-type pulse with the zero mean value (b).}
\label{f05} %
\end{figure}

According to the inverse scattering method \cite{Whth, Karp,
AblSeg}, to study the evolution of the initial perturbation within
the KdV Eq.~(\ref{e01}), one have to solve the complementary
Schr\"odinger equation: \\
\be
\label{e19} %
\psi''(\xi) + \frac{\sigma^2}{6}\left[\vp(\xi) +
E_n\right]\psi(\xi) = 0, %
\ee %
where $\psi$ is the auxiliary function; $\vp$ is the dimensionless
function describing the shape of the initial perturbation (see
above), it has the unit amplitude, i.e., $\vp_{max} = 1$, and the
unit characteristic duration in the dimensionless variables
(\ref{e05}). For the case a) in Fig.~\ref{f05}, $\vp = 1$ at $0
\le \xi \le 1$ and zero beyond this interval. Similarly, for the
case b) in Fig.~\ref{f05}, $\vp = 1$ at $0 \le \xi \le 1$, $\vp =
-1$ at $1 < \xi \le 2$ and zero beyond these intervals. Further,
$\sigma^2$ is the Ursel parameter as defined in Eq.~(\ref{e08})
with $U \equiv U_0$ and $T_p \equiv T$; $E_n < 0$ with $n = 1,\,
2\, \ldots, N$ are eigenvalues of the Schr\"odinger
Eq.~(\ref{e19}). The amplitudes of emerging solitons, $A_n$, are
related with the eigenvalues $E_n$ by the simple relation: \be
\label{e20} %
A_n = -2U_0E_n. %
\ee

The analytical solution to the Schr\"odinger Eq.~(\ref{e19}) with
the rectangular potential function can be readily constructed (see
e.g., \cite{LandLif,Flugge}). Omitting simple, but tedious
calculations, the result can be presented in the form of two
transcendental equations determining eigenvalues $E_n$ of the
Schr\"odinger Eq.~(\ref{e19}):
\be %
\label{e21}%
\tan\left[\sigma^2\sqrt{\frac{\beta}{24\alpha}(1 + E)}\,\right] =
\sqrt{\frac{-E}{1 + E}},\phantom{-} %
\ee
\be%
\label{e22}%
\tan\left[\sigma^2\sqrt{\frac{\beta}{24\alpha}(1 + E)}\,\right] =
-\sqrt{\frac{1 + E}{-E}}. %
\ee

In a similar manner the analytical solution can be found for the
initial perturbation shown in Fig.~\ref{f05}b. The outcome is
presented by the following transcendental equation:
\begin{equation}%
\label{e23} %
\tan\left[\sigma^2\sqrt{\frac{\beta}{6\alpha}(1 + E)}\,\right] =
-\frac{\sqrt{1 - E^2}\left(\sqrt{-E} + \sqrt{1 + E}\right) +
\sqrt{1 + E}\left[\sqrt{-E(1 + E)} + 1 - E\right]\mbox{Th}}
{\sqrt{1 - E^2}\left(\sqrt{-E} - \sqrt{1 + E}\right) +
\left[\sqrt{-E(1 - E)} - \sqrt{1 + E}(1 + E)\right]\mbox{Th}},
\end{equation}
where $\ds \mbox{Th} =
\ds\tanh{\left(\sigma^2\sqrt{\frac{\beta}{6\alpha}(1 -
E)}\right)}$.
\begin{figure}[b!]%
%\vspace*{-8cm}%
\centerline{\includegraphics[width=12cm]{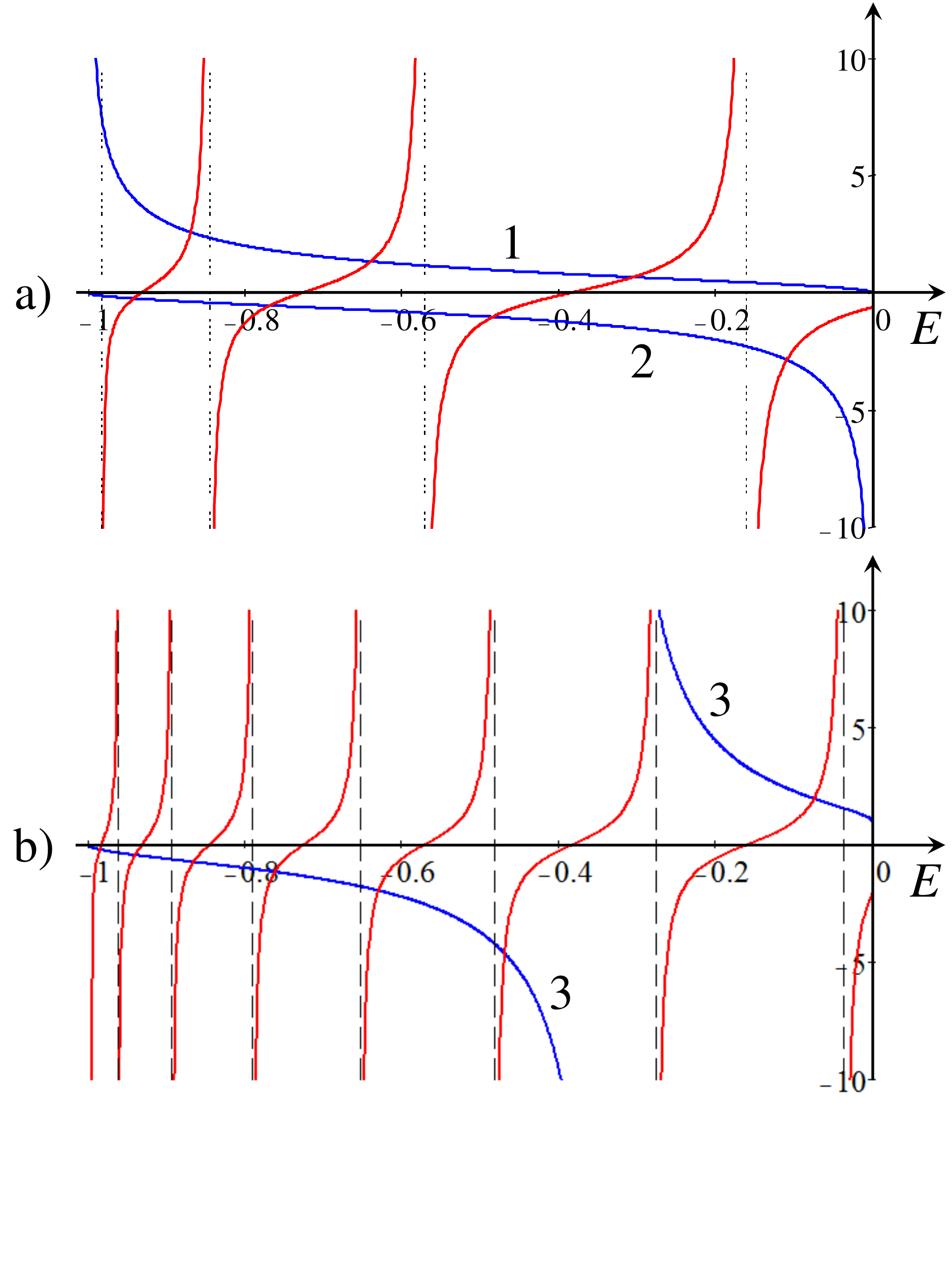}} %
\vspace*{-2.5cm}%
\caption{(color online) (a) Graphical solutions of transcendental
Eqs. (\protect\ref{e21}) and (\protect\ref{e22}). (b) Graphical
solutions of transcendental Eq. (\protect\ref{e23}).} %
\label{f06} %
\end{figure}%

Solutions of the transcendental Eqs. (\ref{e21})--(\ref{e23}) can
be presented graphically as shown in Figs.~\ref{f06}a) and
\ref{f06}b), where the series of tan-type curves represent the
tangential functions in the left-hand side of these equations,
while line 1 represent1 the right-hand side of Eq.~(\ref{e21}),
line 2 represents the right-hand side of Eq.~(\ref{e22}), and two
branches of line 3 represent the right-hand side of
Eqs.~(\ref{e23}).

The plots were generated for the same values of Ursel parameter
$\sigma^2 = 555$, the water depth $h = 1$ m, and the amplitude of
initial perturbations $U_0 = 0.01$ m were chosen the same as in
Subsection 2.1 for the sinusoidal initial functions. In both cases
of rectangular and meander-type pulses the number of roots of
transcendetal equations (\ref{e21})--(\ref{e23}) are the same, $N
= 8$. Dashed vertical lines in each figure show positions of
several first roots of corresponding equations. The amplitudes of
solitons related to these eigenvalues as per Eq.~(\ref{e20}) are
presented in Table 2. \\

Table 2. Amplitudes of solitons emerging from the rectangular and
meander-type initial perturbations shown in Fig.~\ref{f05}.

\smallskip

{\hspace{5mm} %
\begin{tabular}{|l|c|c|c|c|c|c|c|c|}
\hline %
\phantom{wwwww}$A_k \to$ & $A_1$ & $A_2$ & $A_3$ & $A_4$ & $A_5$ & $A_6$ & $A_7$ & $A_8$ \\
Initial pert. $\downarrow$&(mm)&(mm)&(mm)&(mm)&(mm)&(mm)&(mm)&(mm) \\
\hline %
Rectangular pulse & $19.708$ & $18.834$ & $17.382$ & $15.358$ &
$12.780$ & $9.672$ & $6.084$ & $2.168$ \\
\hline %
Meander-type pulse & $19.702$ & $18.806$ & $17.318$ & $15.242$ &
$12.590$ & $9.376$ & $5.638$ & $1.488$ \\
\hline
$RD\cdot 100\%$ & $0.03$ & $0.14$ & $0.36$ & $0.76$ & $1.49$ & $3.06$ & $7.33$ & $31.36$ \\
\hline %
\end{tabular}}

\bigskip

The last row in the Table shows the relative difference in
percents between the corresponding soliton amplitudes for the
rectangular $A^r_k$ and meander-type $A^m_k$ perturbations: $RD =
(A^r_k - A^m_k)/A^r_k$. As one can see from this Table, the
difference between the corresponding amplitudes $A_k$ is fairly
small especially for the first largest solitons, and only for two
last solitons of very small amplitudes the difference amounts for
about 7\% and 31\% correspondingly. It can be readily shown that
the larger the Ursel parameter, the greater the number of emerging
solitons and the smaller the difference in theirs amplitudes.

Thus, one can conclude that the asymptotic theory developed in
\cite{Karp} can provide a good basis for the calculation of
statistical properties of solitons obscured in the random wave
field. Apparently, the most energetic part of the soliton spectrum
(a right wing of soliton distribution function at large
amplitudes) is described fairly good while at small amplitudes the
distribution function may be not quite correct.
\begin{figure}[h!]
\vspace*{-10cm}%
\centerline{\includegraphics[width=16cm]{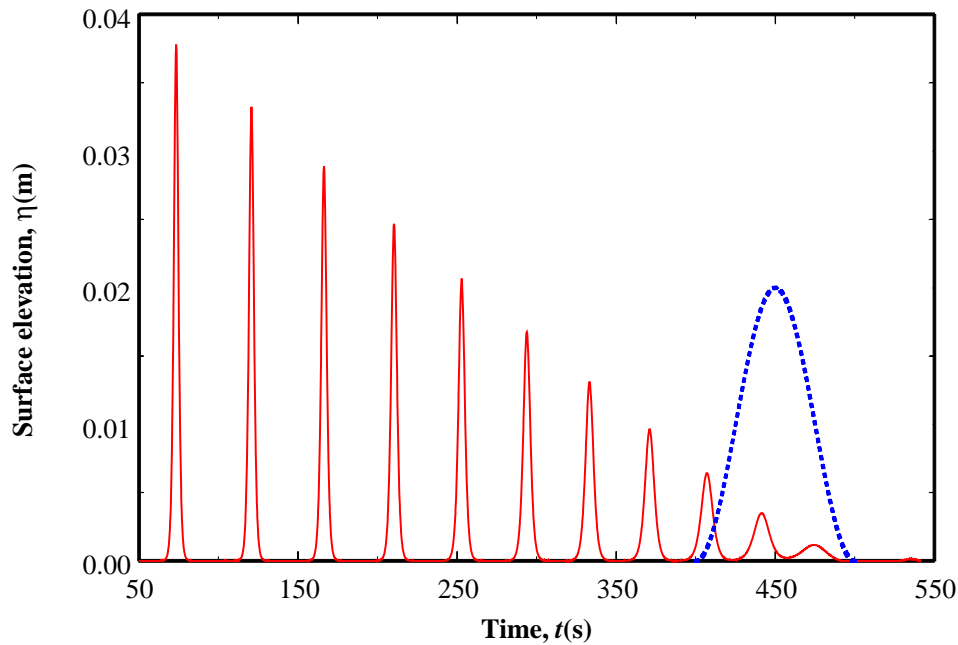}} %
\vspace*{-2cm}%
\caption{(color online) Cosine initial pulse (dashed line) and the
result of its evolution at the distance $x = 6\cdot10^4$ m (solid line).}%
\label{f07} %
\end{figure}

\subsection{Cosine initial pulse}

Let us now consider why the number of solitons emerging in each
period of a pure periodic perturbation (as well as their
amplitudes) differs from those emerging from the pulse-type
perturbations of the same shape, amplitude and duration (cf.
Fig.~\ref{f02} with Figs.~\ref{f03} and \ref{f04})? The answer is,
apparently, as follows: In the pure periodic case the zero level
of the physical system is not determined clearly. The system
admits {\it the minimum of the wave field} as the zero level.
Hence, a periodic sinusoidal perturbation can be treated as a
periodic sequence of positive cosine-type pulses with respect to
the minimum possible level. In a process of evolution, each pulse
disintegrates into a number of solitons whose total amount is much
greater than for sine-type perturbations considered above. New
numerical simulation was carried out with this modified zero level
for single cosine-type pulse as the initial perturbation; the
result is shown in Fig.~\ref{f07}.

There were clearly detected 11 solitons whose amplitudes are
presented in Table 3. The amplitudes of solitons emerging from the
periodic sinusoidal perturbation described above (see Subsection
\ref{Subsec2.1.1}) are also shown in the same Table for the
comparison. This time soliton amplitudes for the sinusoidal
perturbation were calculated more thoroughly using trace method
suggested in \cite{SalMauEng}. When the sinusoidal perturbation
disintegrates for the first time into the sequence of solitons
ordered by their amplitudes as shown in Fig.~\ref{f02}, the
observer can not see all solitons because some of them are still
obscured. We followed up for the development of the wave field
after the distance $x = 1.1\cdot 10^4$ m which corresponds to
Fig.~\ref{f02} and discovered that in the process of soliton
interactions some more solitons of small amplitudes appear. So
that the total number of solitons in this case was exactly the
same as for the cosine initial pulse, $N = 11$. Their amplitudes
were measured with respect to the minima of the initial sinusoidal
perturbation. As one can see, the relative difference in percents
between the corresponding soliton amplitudes for the cosine
initial pulse and periodic sinusoidal perturbation, $RD = (A^c_k -
A^s_k)/A^c_k$ is not too big now and not exceeds 17\%. (see the
last row of Table 3). \\

Table 3. Amplitudes of solitons, $A_k$, emerging from the cosine
initial pulse and from the periodic sinusoidal perturbation.

\smallskip

{\hspace{5mm} %
\begin{tabular}{|l|c|c|c|c|c|c|c|c|c|c|c|}%
\hline %
\phantom{wwwww}$A_k \to$ & $A_1$ & $A_2$ & $A_3$ & $A_4$ & $A_5$
 & $A_6$ & $A_7$ & $A_8$ & $A_9$ & $A_{10}$ & $A_{11}$ \\
Initial pert. $\downarrow$&(mm)&(mm)&(mm)&(mm)&(mm)&(mm)&(mm)&(mm)&(mm)&(mm)&(mm) \\
\hline %
Cos., pulse & $37.75$ & $33.25$ & $28.89$ & $24.68$ & $20.63$ &
$16.77$ & $13.09$ & $9.636$ & $6.422$ & $3.49$ & $1.15$ \\
\hline Sin., periodic & $33.70$ & $30.39$ & $26.92$ & $23.36$ &
$19.60$ &
$15.54$ & $11.43$ & $8.04$ & $5.66$ & $-$ & $-$ \\
\hline %
$RD\cdot 100\%$ & $10.73$ & $8.60$ & $6.82$ & $5.35$ & $4.98$ &
$7.35$ & $12.65$ & $16.60$ & $11.82$ & $-$ & $-$ \\
\hline %
\end{tabular}}

\bigskip

Thus, if the zero level of the considered physical system is known
and the perturbation eventually vanishes at the infinity then the
statistics of obscured solitons is determined by positive humps of
the perturbation with respect to this zero level. Namely such a
situation takes place in the case of hydrophysical measurements in
laboratory (water tanks) or in natural marine conditions. But if
the perturbation is periodic in principle, then solitons, their
numbers and amplitudes are determined by each hump of the
perturbation with respect to the total minimum of the initial
perturbation.

\subsection{Data processing. A model example.}
\label{Subsec2.4}

\subsubsection{A model spectrum and the range of its validity}
\label{Subsec2.4.1}

Let us apply now the developed approach to the random perturbation
artificially generated and presented in Fig.~\ref{f01}. This
perturbation was obtained by means of inverse Fourier transform of
a series of 128 harmonics having random phases in the interval [0,
$2\pi$] and amplitudes distributed in accordance with the formula
\be %
\label{e24} %
S(\omega) = S_0(\omega + \omega_0)^{-2}\tanh{\frac{\omega}
{\delta\omega}}, %
\ee %
where $S_0 = 2\cdot 10^{-4}$ m, $\omega_0 = 2\pi\cdot10^{-2}$
s$^{-1}$, and $\delta\omega = 2.5\cdot 10^{-2}$ s$^{-1}$.

\begin{figure}[b!]
\vspace*{-10cm}%
\centerline{\includegraphics[width=16cm]{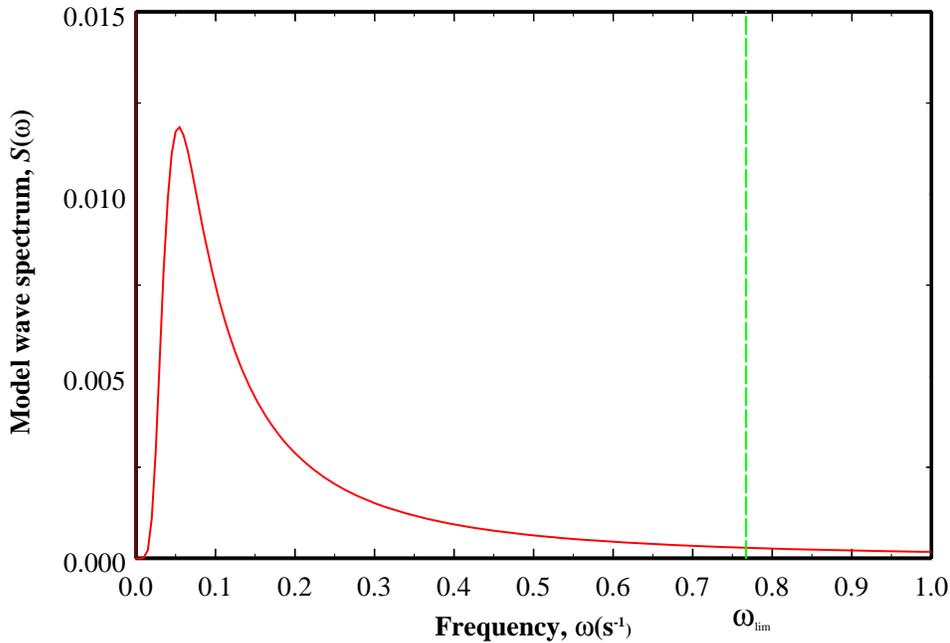}} %
\vspace*{-2cm}%
\caption{(color online) The model amplitude spectrum of random
wave field shown in Fig.~\protect\ref{f01}.}
\label{f08}%
\end{figure}

For the construction of quasi-random wave field it was taken only
the low-frequency and most energetic part of this spectrum, $0 \le
\omega \le \omega_{lim}$, where $\omega_{lim} = 0.767$ s$^{-1}$
(see vertical dashed line in Fig.~\ref{f08}). The limiting
frequency, $\omega_{lim}$, was chosen for the following reasons.
The dispersion relation for infinitesimal perturbations within the
TKdV Eq.~(\ref{e01}) is
\be %
\label{e25} k = \frac{\omega}{c_0} + \frac{h^2}{6c_0^3}\omega^3  =
\frac{\omega}{c_0}\left(1 + \frac{h^2}{6c_0^2}\omega^2\right), %
\ee %
where $\omega$ is the frequency of a sinusoidal wave, and $k$ is
the wave number, $\eta \sim e^{i(\omega t - kx)}$.

As is well known, this dispersion relation represents an
approximation of a real dispersion relation of a physical system,
e.g., water or plasma waves, when $\omega, k \to 0$. The range of
validity of the dispersion relation (\ref{e25}) is restricted by
the requirement that the second term in brackets is small in
comparison with one (see, e.g., \cite{AblSeg,Karp,Whth}). This
condition gives
\be %
\label{e26} %
\omega \ll \omega_{cr}
\equiv \frac{c_0}{h}\sqrt{6} = \sqrt{\frac{6g}{h}}. %
\ee

In our case $\omega_{cr} \approx 7.67$ s$^{-1}$. To satisfy
condition (\ref{e26}), we set $\omega_{lim} = 0.1\omega_{cr} =
0.767$ s$^{-1}$.

\subsubsection{Data processing of the model initial perturbation}
\label{Subsec2.4.2}

\begin{figure}[ht] %
\vspace*{-10cm}%
\centerline{\includegraphics[width=16cm]{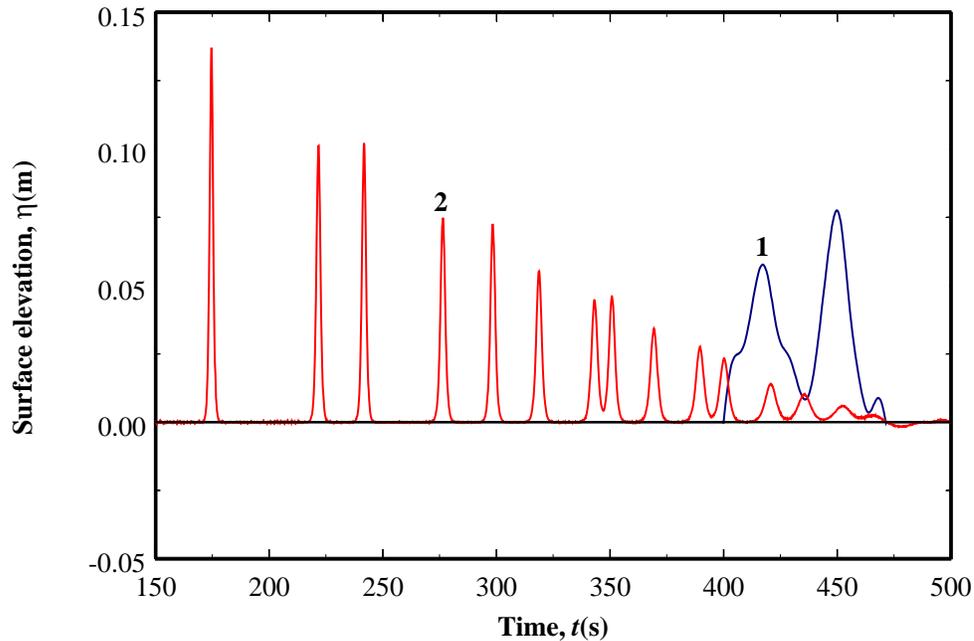}} %
\vspace*{-2cm}%
\caption{(color online) Disintegration of one of the initial
random pulses (line 1) on solitons (line 2).} %
\label{f09} %
\end{figure}

Let us consider now a quasi-random wave field shown in
Fig.~\ref{f01} and apply the approach developed above. Each
positive hump enumerated in the figure can be studied separately
by means of the numerical code for the TKdV Eq.~(\ref{e01}). The
numerical code was based on the explicit finite-difference scheme
of a second-order accuracy both on spatial and temporal variables
\cite{Berez}. The theoretical analysis shows that used
central-difference scheme is conditionally stable provided that
$\Delta x \sim \Delta t^3$, where $\Delta x$ and $\Delta t$ are
the spatial and temporal mesh steps. The code works very
effectively and fast so that the result of pulse fission on
solitons was obtained very quickly. A typical picture of
disintegration of one of the pulses (pulse No 5 in Fig.~\ref{f01})
is shown in Fig.~\ref{f09}.

\begin{figure}[ht!]%
\vspace*{-9cm}%
\centerline{\includegraphics[width=15cm]{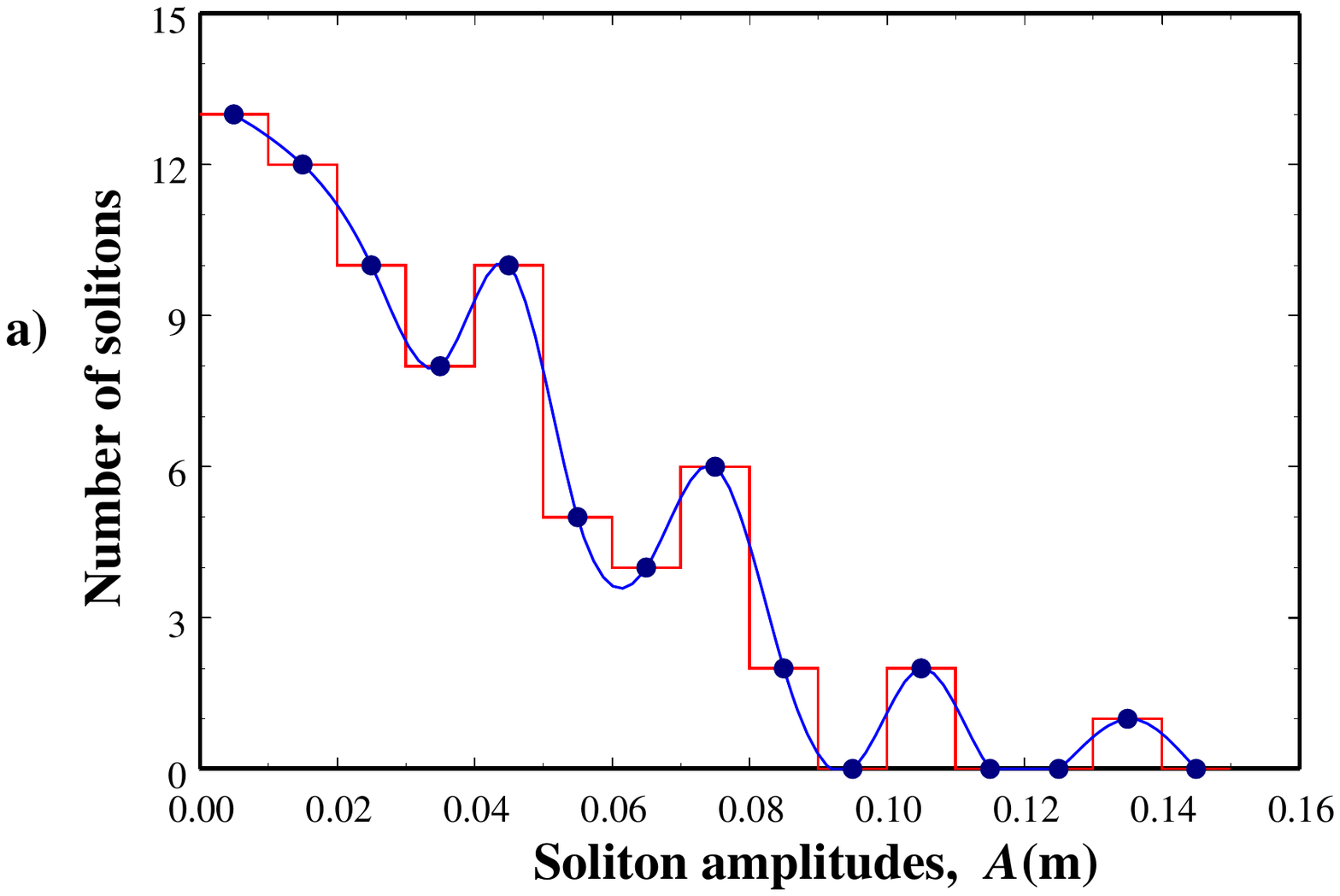}} %
\vspace*{-11cm}%
\centerline{\includegraphics[width=15cm]{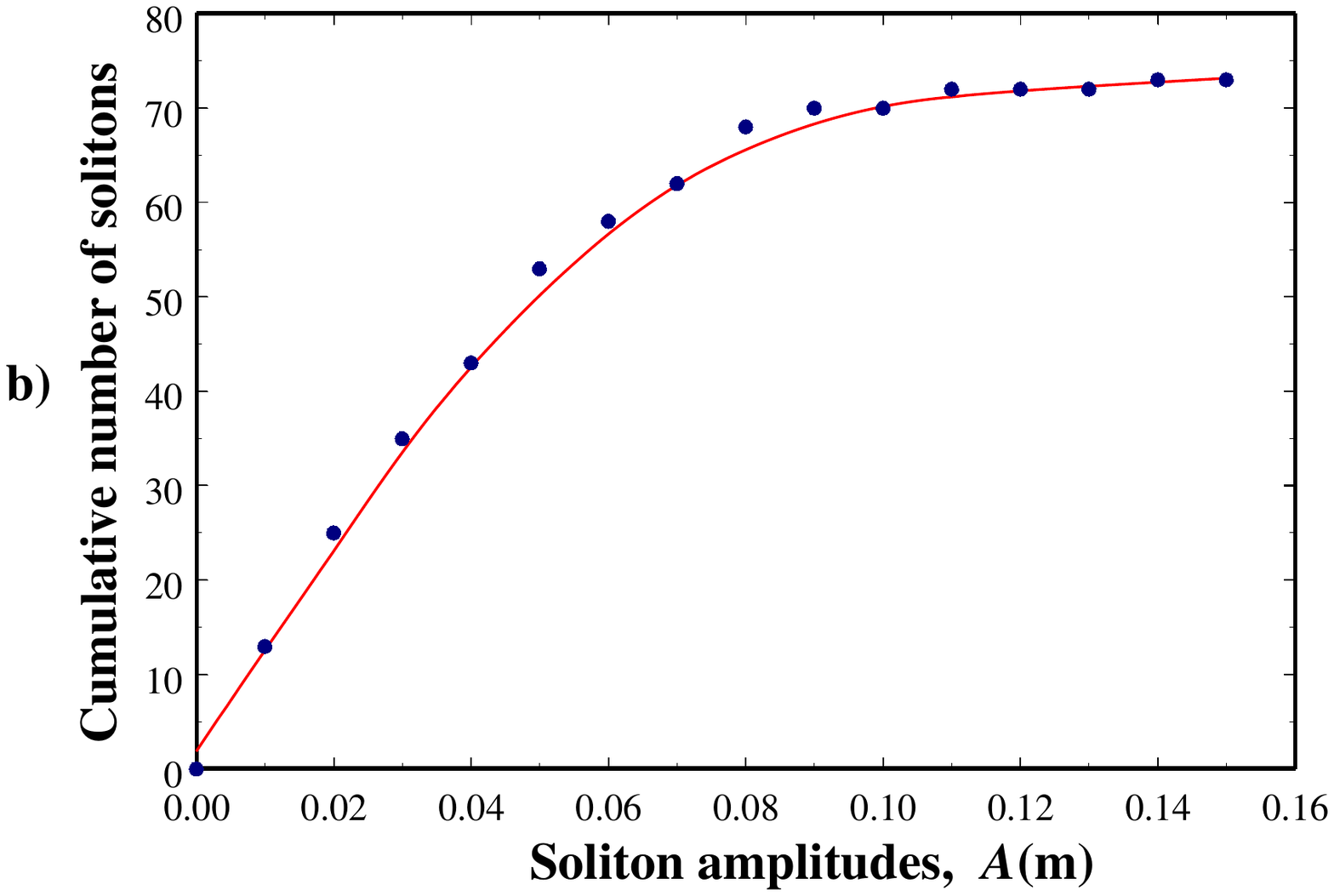}} %
\vspace*{-1.5cm}%
\caption{(color online) Frame a) -- the histogram of soliton
distribution (piece-linear line) and the model distribution
function (smooth line) for the illustrative example. Frame b) --
the cumulative distribution function versus soliton amplitude
corresponding to the model distribution function shown in frame
a). Smooth lines in both frames were obtained by means of spline
interpolation of numerical data.}
\label{f10} %
\end{figure}

The number of emerged solitons and their parameters can be easily
calculated. To avoid errors in determination of soliton
parameters, the numerical calculations were conducted until each
soliton was sufficiently separated from others, so that their
fields were not overlapped in the vicinity of their maxima. This
procedure was carried out for each pulse shown in Fig.~\ref{f01}).
As the result, it was obtained a large number of solitons of
different amplitudes, their total number for all pulses was $N_s =
73$ which is enough for the illustrative purposes. The solitons
were collected into several groups (15 groups in total) according
to their amplitudes. This allowed us to construct a histogram of
numbers of solitons versus their amplitudes. The histogram can be
considered as a model of the distribution function $f(A)$ of
density of number of solitons on amplitudes (see Eq.~(\ref{e07})).
The histogram obtained and the model distribution function for the
considered illustrative example are shown in Fig.~\ref{f10}. Frame
b) in the figure shows the cumulative distribution function, i.e.,
the total number of solitons whose amplitudes are not greater than
the given value.

\section{Laboratory experiment with wind waves on shallow water}
\label{Subsec2.5}

The theory developed above was applied to the data of laboratory
experiments with wind wave generation. Series of experiments were
carried out in the Luminy (Marseilles) small tank made of
plexiglass and having the following sizes (length $\times$ width
$\times$ height): 865 cm $\times$ 64 cm $\times$ 50 cm. The water
depth in the tank in different experiments ranged from 1 cm to 8
cm: $h = 1,\;2,\;3,\;4,\;6,$ and 8 cm. Surface waves were
generated by a wind blowing up over the water surface with the
different mean velocities: $V_w = 5.29, \; 6.45,\;8.62$ and 13.24
m/s. The ventilator producing an air flow was installed at one of
the ends of the tank. The blower width was the same as the width
of the tank, 64 cm, but its height was 31 cm above the water
level. The tank was covered above the blower by a plexiglass lid.
At the opposite end of the tank it was placed a wave absorber to
exclude reflected waves (an inclined bottom causing surface wave
breaking). Two sensitive electric probes recording water level
were placed at the distances 100 cm and 300 cm from the
ventilator. The probes were thoroughly calibrated before each
experiment; their sensitivities were equal and amounted $s = 0.61$
V/cm.

Below we present the analysis of only one of the series of
experiments with the water depth $h = 1$ cm and wind velocity $V_w
= 5.29$ m/s. Other experimental series were analysed in a similar
way. We have to make a reservation in advance that the
experimental data with wind waves are not perfect for the
illustration of suggested approach. Wind waves generated by
permanently blowing wind is an active system, i.e., the system
with the permanent energy pumping at each point of water surface.
Moreover, a distributed external force due to wind is not a
constant, it varies from some maximum value near the ventilator to
some small value at the opposite end of the tank. This results in
the different soliton distribution functions measured at two
different distances from the ventilator. A small water viscosity
can also affect the soliton distribution function.
\begin{figure}[h!]
\vspace*{-3cm}%
\centerline{\includegraphics[width=21cm]{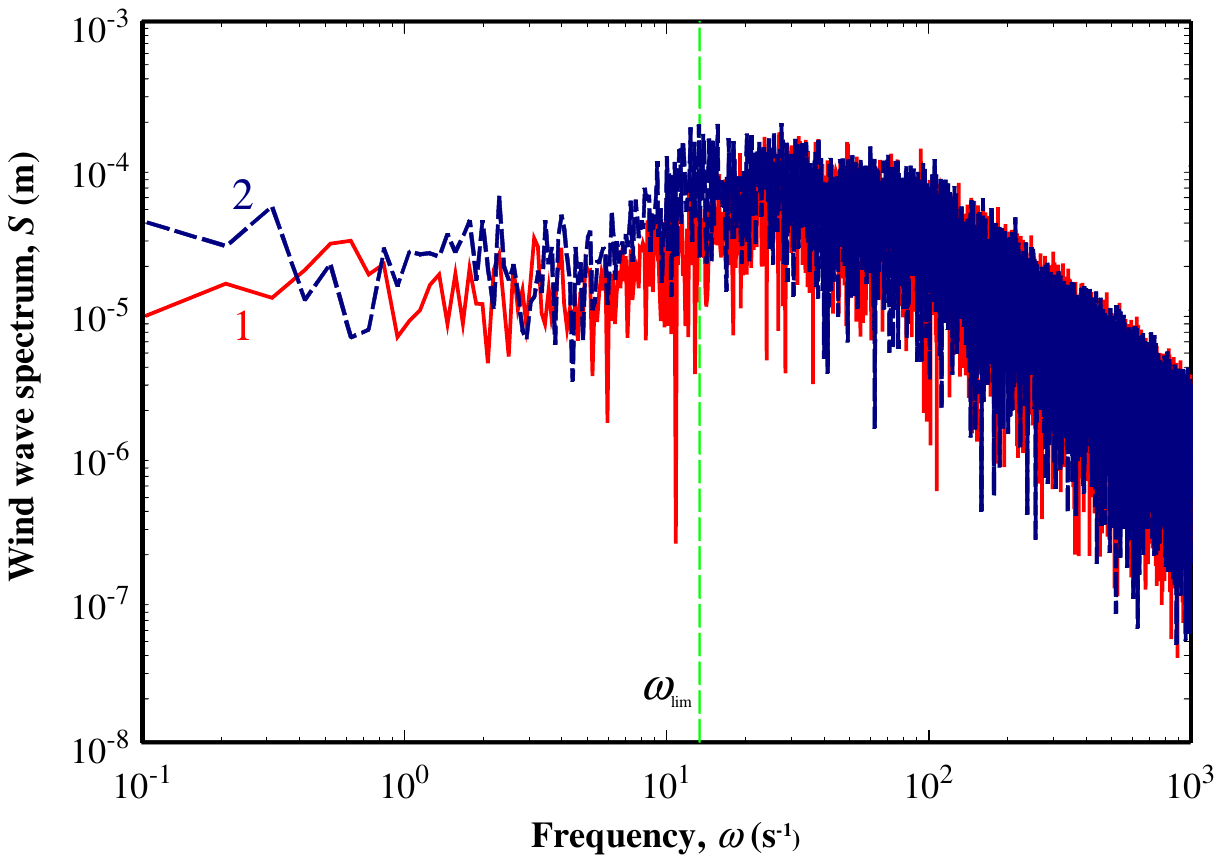}} %
\vspace*{-18cm}%
\caption{(color online) Fourier spectrum of wind waves generated
in the laboratory tank at two different distances from the
ventilator. Solid line 1 represents the spectrum at the distance
100 cm; dashed line 2 represents the spectrum at the
distance 300 cm.}%
\label{f11} %
\end{figure}

Another difficulty with surface waves generated by wind is the
effective generation of high frequency Fourier components, so that
the essential portion of wave energy is contained in that part of
Fourier spectrum which is beyond the range of validity of KdV
equation. Therefore we were forced to restrict our analysis by
only the low-frequency components of the wave spectrum. Figure
\ref{f11} presents the Fourier spectrum of wind waves recorded at
two distances from the ventilator as indicated above. The critical
frequency at $h = 1$ cm is $\omega_{cr} = 76.7$ s$^{-1}$, and to
satisfy the condition (\ref{e26}) we cut the Fourier spectra of
wind waves at $\omega_{lim} = 13.4$ s$^{-1}$ (see dashed vertical
line in Fig.~\ref{f11}) and ignored the high frequency portions of
spectra above $\omega_{lim}$.

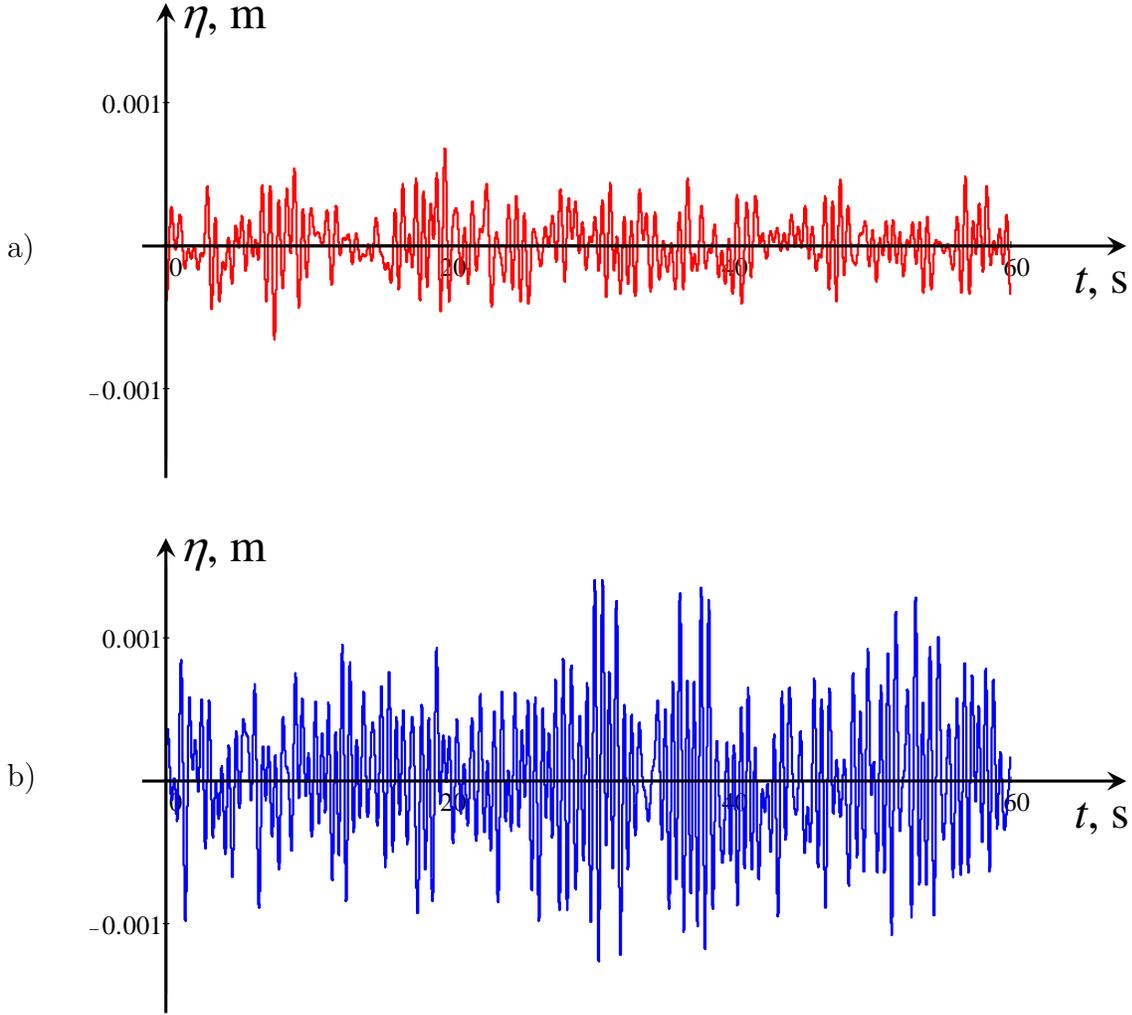
\begin{figure}[h] %
%\vspace*{-10cm}%
\centerline{\includegraphics[width=14cm]{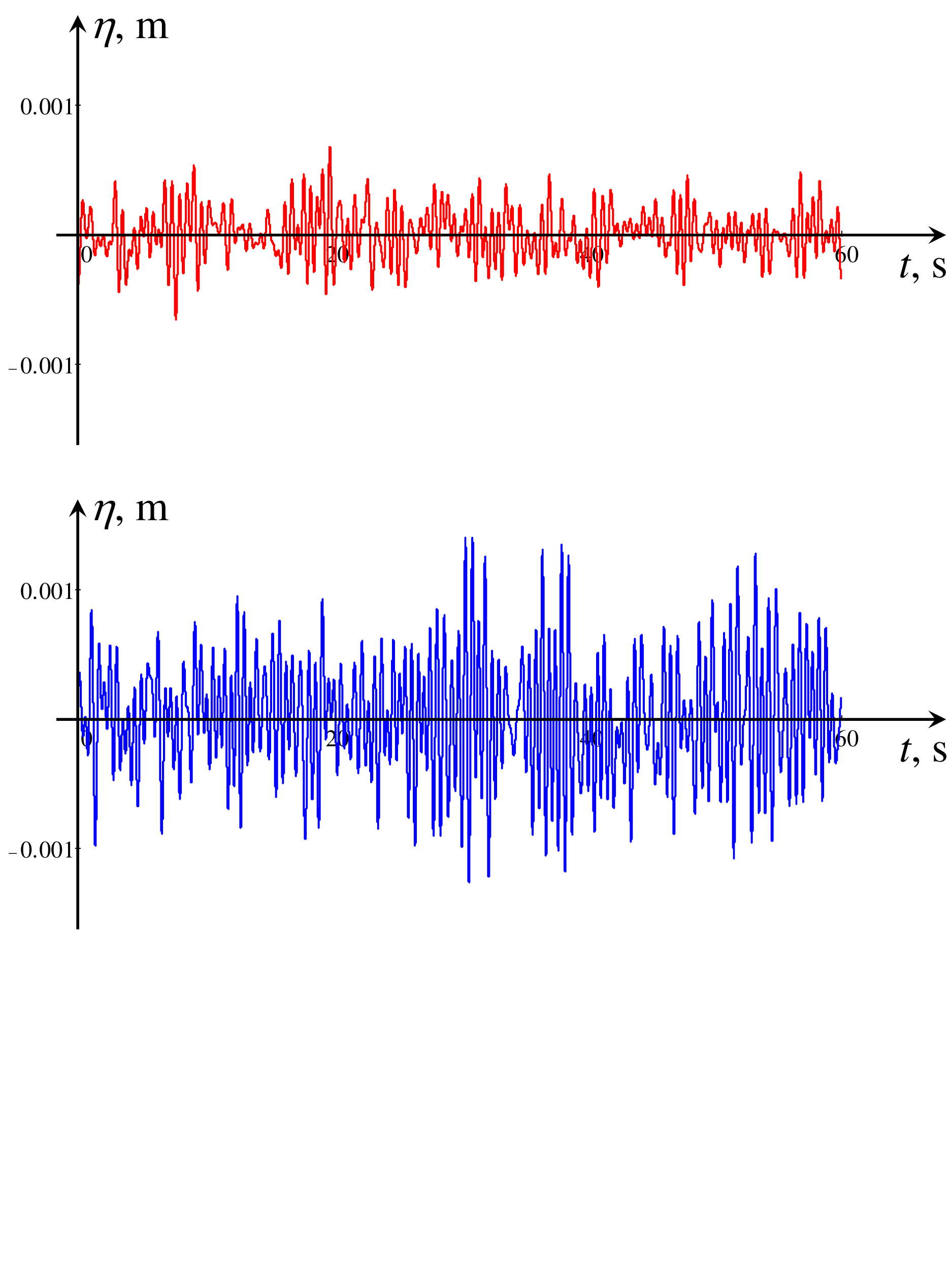}} %
\begin{picture}(300,6)%
\put(-75,440){{\large a)}}%
\put(-75,240){{\large b)}}%
\end{picture}
\vspace*{-4.5cm}%
\caption{(color online) Fragments of low-frequency components of
surface perturbations generated by wind in two spatial points of
measurements in the tank. Frame a) -- surface perturbation at the
distance 100 cm; frame b) -- at the distance 300 cm.}
\label{f12} %
\end{figure}

The surface perturbation was reconstructed by means of the inverse
Fourier transform on the basis of extracted portions of wave
spectra in the range $0 < \omega \le \omega_{lim})$. The fragments
of 60-second duration records corresponding to two spatial points
of measurements are shown in Fig.~\ref{f12}.

As it is clearly seen from the comparison of two data records
presented in frames a) and b), there is the essential difference
between them, especially in the intensity of wave fields. This can
be explained by the effect of a fetch on wind wave generation.
Therefore, one can expect that the statistics of obscured solitons
in the record of frame b) is essentially richer than in frame a).

The recorded data shown in Fig.~\ref{f12} were used as the input
data for the TKdV equation (\ref{e01}). In the statistically
equilibrium state each 60-second portion of recorded data is
equivalent to the same portion taken at a different time,
therefore one can expect that the number of solitons obscured in
each portion of data is the same in average and their distribution
function is invariant with respect to time shift. This was
confirmed in the data processing.

\begin{figure}[h!]%
\centerline{\includegraphics[width=18cm]{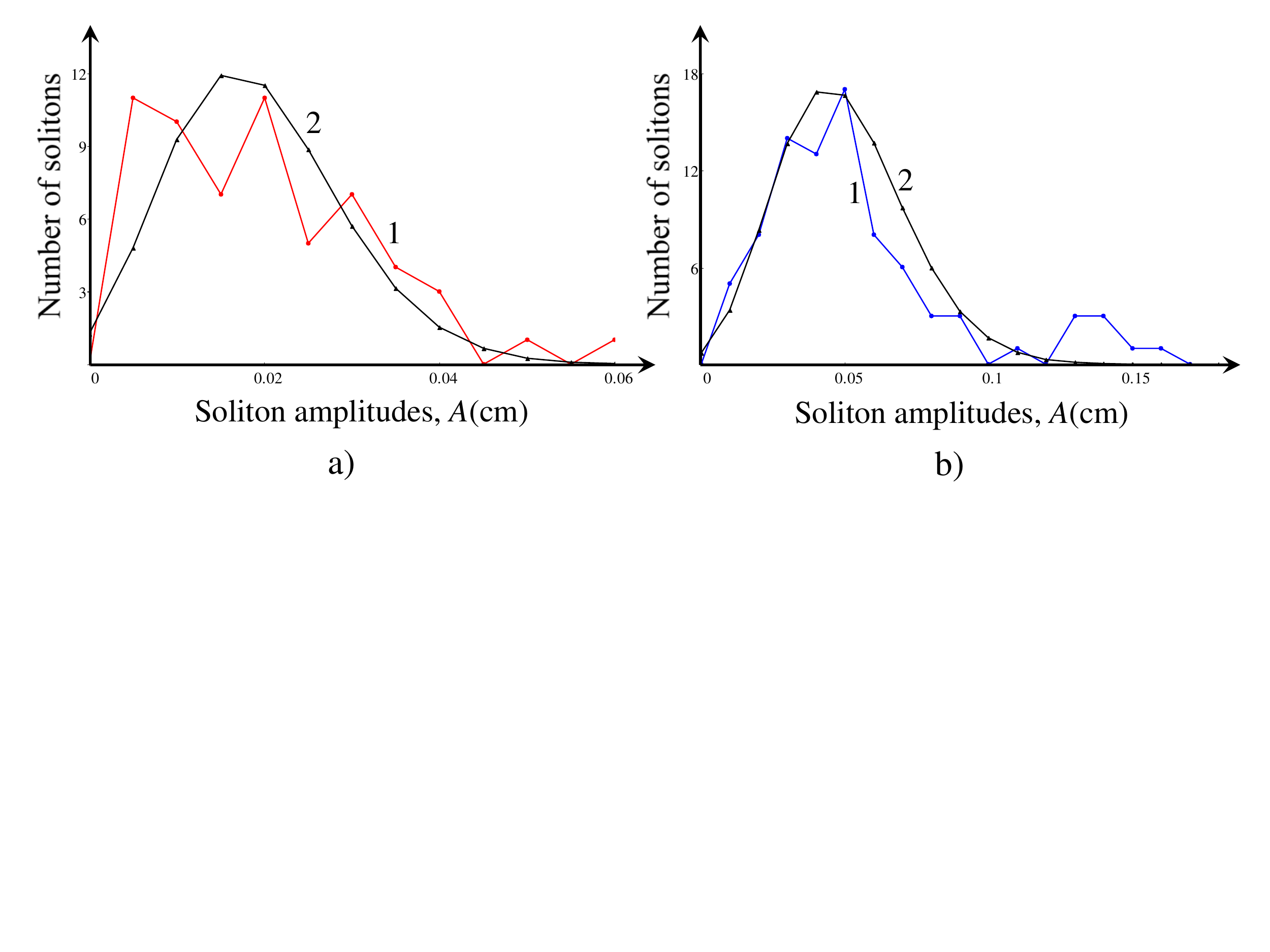}} %
\vspace*{-6.5cm}%
\caption{(color online) The histogram of soliton numbers versus
soliton amplitudes for the time series shown in Fig.~\ref{f12}.
Line 1 in each frame represent experimental data, and lines 2 --
the best fit of these data by the Poisson distribution functions
with the parameters $\lambda = 3.85$ for the histogram shown in
frame (a) and $\lambda = 4.94$ for the histogram shown in frame
(b).}
\label{f13} %
\end{figure}

The TKdV equation (\ref{e01}) was solved numerically using the
recorded data of 60 sec duration from the total time interval of
208 sec. After a while solitons emerged from the quasi-random
data, and their amplitudes were easily determined with the help of
a special subroutine. This allowed us to determine the histogram
of soliton numbers in the each particular interval of amplitudes
$A + \Delta A$; this is the analog of a differential distribution
function. On the basis of this function we determined also the
integral (cumulative) distribution function -- the total number of
solitons with the amplitudes less than $A$ normalised by the total
number of all solitons. Figure \ref{f13} demonstrates the
histogram of soliton numbers versus amplitudes for the time series
shown in Fig.~\ref{f12}. The experimental data can be approximated
by the Poisson distribution function $P(n) =
\lambda^ne^{-\lambda}/n!$, where the parameter $\lambda = 3.85$
for the histogram shown in frame (a) and $\lambda = 4.94$ for the
histogram shown in frame (b).
\begin{figure}[ht!]%
\centerline{\includegraphics[width=18cm]{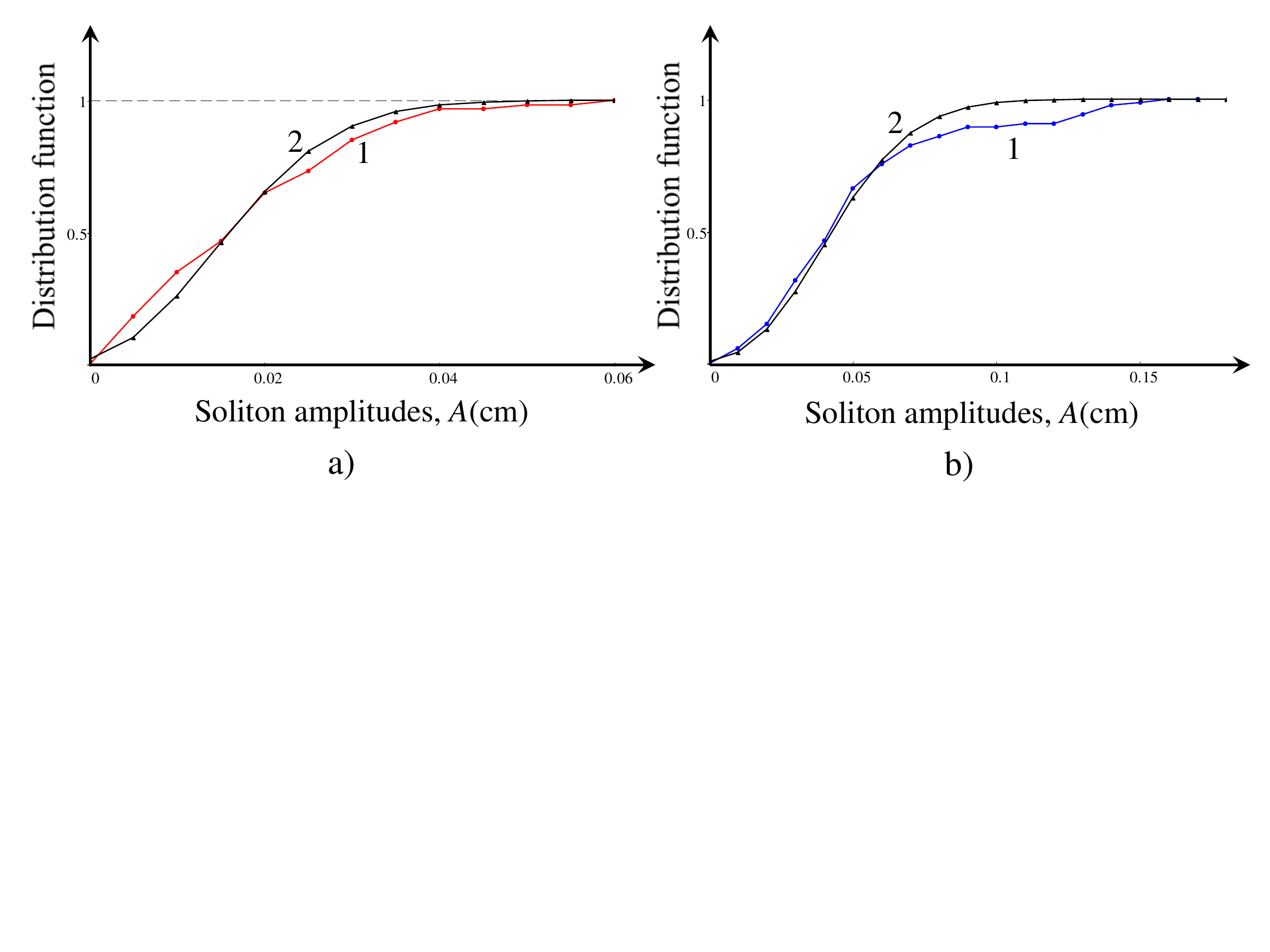}} %
\vspace*{-6.5cm}%
\caption{(color online) The integral (cumulative) distribution
functions (lines) for the experimental data shown in
Fig.~\ref{f12}. Lines 2 represent the best fit approximation with
the Poisson cumulative distribution functions with the same
parameters $\lambda$ as in Fig.~\ref{f13}.}
\label{f14} %
\end{figure}

The corresponding integral distribution functions for the
experimental data of Fig.~\ref{f12} are shown in Fig.~\ref{f14}
(lines 1) together with the approximative Poisson integral
functions with the same parameters as in Fig.~\ref{f13}. The total
number of solitons emerged from the wave field shown in
Fig.~\ref{f12}a) was 60, and emerged from the wave field shown in
Fig.~\ref{f12}b) was 86. As expected, the time series recorded
closer to the ventilator (Fig.~\ref{f12}a) contained less number
of solitons than the time series recorded further from the
ventilator (Fig.~\ref{f12}b). In the latter case the wave field
was much better developed due to the influence of wave fetch.

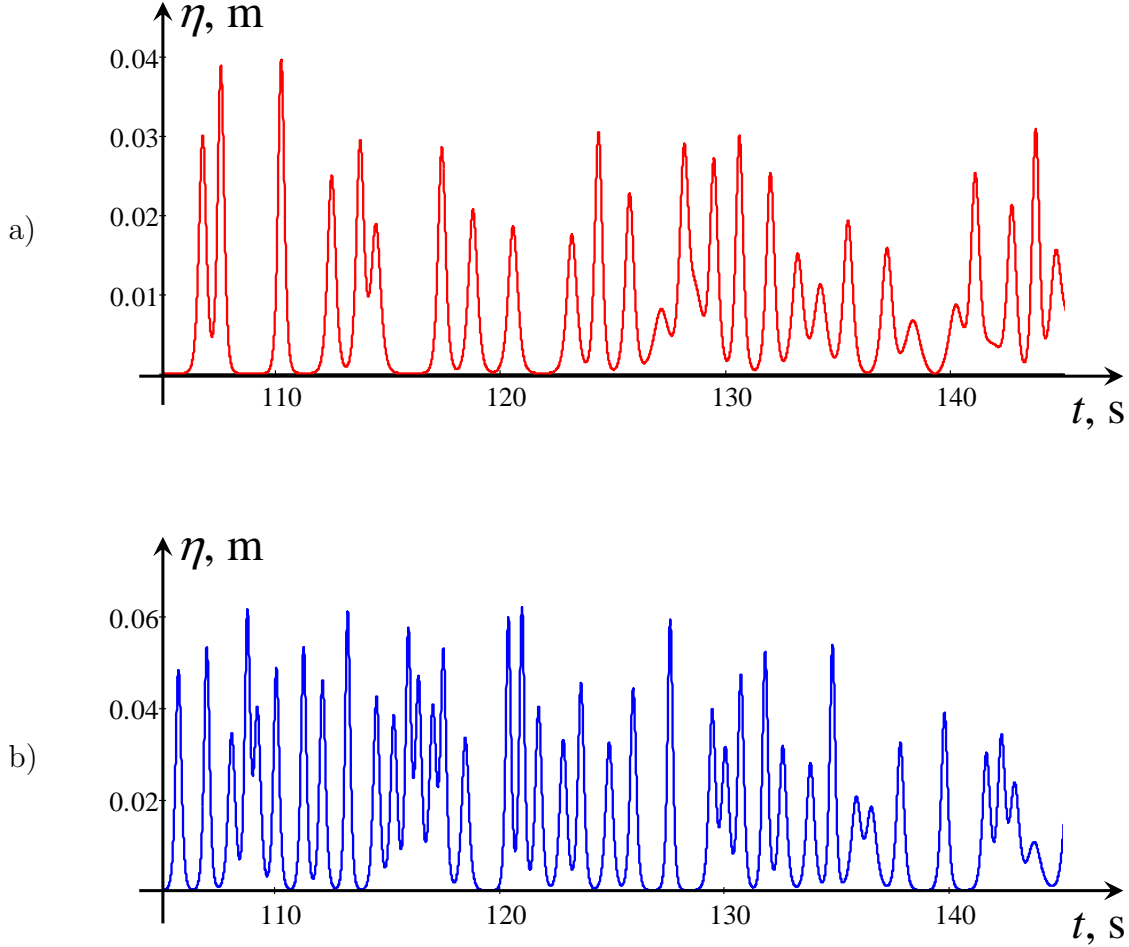
\begin{figure}[ht!]%
\centerline{\includegraphics[width=14cm]{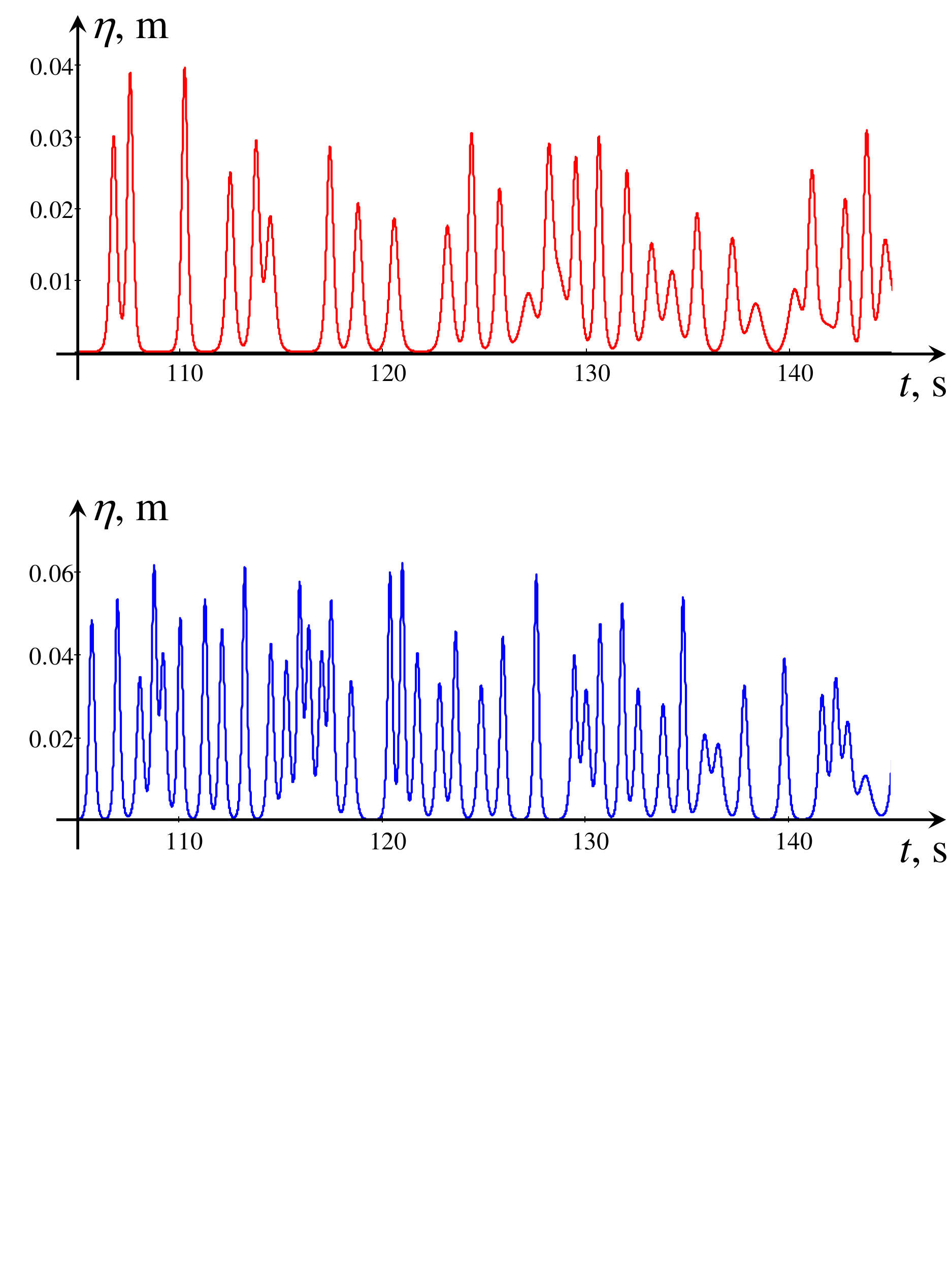}} %
\vspace*{-5.5cm}%
\begin{picture}(300,6)%
\put(-75,290){{\large a)}}%
\put(-75,90){{\large b)}}%
\end{picture}
\caption{(color online) The fragments of numerical calculations
with the input data taken from Fig.~\ref{f12} illustrating the
soliton gas density in both time series.}
\label{f15} %
\end{figure}

If we assume that all 60 solitons in the time series shown if
Fig.~\ref{f13}a) are uniformly distributed in the time interval of
60 s, then we obtain that the time interval per each soliton is
$\Delta T_1 = 1$ s. As follows from the histogram shown in
Fig.~\ref{f13}a) the maximal number of solitons have amplitudes
$A_{m1} = 0.02$ cm and the duration $T_{s1} = 0.026$ s. Therefore
$\Delta T_1/T_{s1} \approx 3.85$. The similar estimates for the
time series shown if Fig.~\ref{f13}b) give the time interval per
each soliton $\Delta T_2 \approx 0.7$ s. As follows from the
histogram shown in Fig.~\ref{f13}b) the maximal number of solitons
in this time series have amplitudes $A_{m2} = 0.05$ cm and the
duration $T_{s2} = 0.165$ s. Therefore $\Delta T_2/T_{s2} \approx
4.2$. Thus, we see that in both cases the ``soliton gas'' is very
dense (cf. \cite{El-16, Shurgalina}). The fragments of numerical
calculations presented in Fig.~\ref{f15} illustrate the soliton
gas density in both time series.

\section{Conclusion}

To analyze long random time series of water waves in shallow
basins we have proposed an approach which differs from the
traditionally used Fourier analysis. Our approach is based on the
extraction of obscured solitons from the complex wave fields and
construction of histograms of solitons at different points of
observation. The histograms can be considered as experimental
counterpart of distribution functions of number of solitons on
their amplitudes. According to the theoretical conception, a
soliton component of a wave field in the well-developed nonlinear
perturbations should dominate. As is well known, the number and
individual parameters of solitons preserve in the conservative
statistically homogeneous systems \cite{Zakh-71, Zakh-09},
therefore the distribution function (or histograms of solitons) is
the same at different points of observation if the dissipative
factors (i.e., viscosity or external sources of energy) are
negligible. In contrast to that the Fourier spectrum changes due
to nonlinearity.

Our approach is in line with the contemporary development of the
theory of strong turbulence in the integrable or near-integrable
systems \cite{El-03, ElKam-05, Zakh-09, Costa, Dytukh-14,
Shurgalina, El-16, Carbone-16}. Experimentally constructed
distribution function can be used for the determination of degree
of soliton gas density -- how far the density is from the critical
value as defined in Refs. \cite{Shurgalina, El-16}. Data
processing of laboratory experiments presented in our paper
supplement the data processing of field experiments reported in
Ref. \cite{Costa}.

A small dissipation can cause a gradual decay of soliton
histograms and their distortion, in general. The histogram decay
and its distortion depend on the specific type of dissipation;
this problem has not been studied yet, although the decay of
individual solitons under the influence of various types of
dissipation has been investigated for the KdV \cite{Grimshaw-03}
and Benjamin--Ono \cite{Grimshaw-18} solitons, as well as for the
Kadomtsev--Petviashvili lumps \cite{Clarke}.

Our approach can provide some additional valuable information
about the energy distribution in natural wave fields such as the
relationship between the soliton and nonsoliton components of the
perturbation, and may indicate on the existence and intensity of
external sources or sinks of energy.

To determine the soliton number and amplitudes from the random
time series, we applied direct numerical modelling for the
evolution of initial data within the framework of the TKdV
equation (\ref{e01}). The existing numerical codes (see, e.g.,
\cite{Berez}) allow us to obtain the results fairly quickly in the
form convenient to further analysis. However, it is not the only
method which can be applied; the numerical solution of the
eigenvalue problem (\ref{e19}) can also be convenient and useful.
Our experience with the application of the approach developed here
shows that there is no problem with the determination of number
and amplitudes of intense solitons, i.e., solitons of big and
moderate amplitudes. However, it takes more efforts to determine
parameters of solitons whose amplitudes are very small. In
general, an uncertainty in the determination of parameters of
solitons of very small amplitudes is higher than of moderate and
big solitons. Meanwhile, the model example of subsection
\ref{Subsec2.4.2} shown in Fig.~\ref{f10}, as well as the
experimental laboratory data shown in Fig.~\ref{f13} demonstarte
that the number of such small-amplitude solitons may be relatively
big.

As has been mentioned, the soliton distribution function remains
unchanged in the integrable systems. However, the wave field
randomly fluctuates in the process of evolution. This leads to the
random fluctuations of local wave extrema. As has been shown in
Ref. \cite{Shurgalina}, the distribution function of wave extrema
varies with time even when the wave field consists of solitons
only. This can be of interest from the viewpoint of physical
applications, but beyond the scope of this paper.

In conclusion, we emphasize that the approach developed here is
applicable to the KdV-type systems, e.g., shallow-water waves
(see, for example, Ref. \cite{Dytukh-14} where the turbulence of
soliton gas was studied both within the integrable KdV and
non-integrable KdV-BBM equations). Its generalization to
deep-water waves described by the Benjamin--Ono or nonlinear
Shr\"odinger equation is the interesting and challenging problem.

\bigskip %

{\bf Acknowledgments.} This work was initiated while one of the
authors (Y.S.) was the invited visitor at the Laboratoire
IRPHE--IOA, Marseille, France several years ago. Y.~S. is grateful
to CNRS, France for the invitation and to the Laboratory staff for
the hospitality. Y.S. also acknowledges the funding of this study
from the State task program in the sphere of scientific activity
of the Ministry of Education and Science of the Russian Federation
(Project No. 5.1246.2017/4.6) and grant of the President of the
Russian Federation for state support of leading scientific schools
of the Russian Federation (NSH-2685.2018.5). The authors are
thankful to Efim Pelinovsky for useful discussions and valuable
advices.

\end{document}